\documentclass[prd,nofootinbib,preprint,superscriptaddress]{revtex4}
\usepackage{amsmath, amssymb, amsthm, graphicx, epsfig, fancyhdr,epsfig, slashed, mathrsfs, comment}
\usepackage{braket}
\usepackage{tikzsymbols}
\usepackage{tikz}
\usepackage{caption}
\usepackage{tikz-feynman}
\tikzfeynmanset{compat=1.1.0}
\usepackage{natbib}
\usepackage{float}
\usepackage{pifont}
\usepackage{booktabs}

\usepackage{tikz,xcolor,hyperref}
\usepackage{subfig}

\definecolor{lime}{HTML}{A6CE39}
\DeclareRobustCommand{\orcidicon}{
	\begin{tikzpicture}
	\draw[lime, fill=lime] (0,0) 
	circle [radius=0.2] 
	node[white] {{\fontfamily{qag}\selectfont \tiny ID}};
	\draw[white, fill=white] (-0.0625,0.095) 
	circle [radius=0.007];
	\end{tikzpicture}
	\hspace{-2mm}
}

\foreach \x in {A, ..., Z}{\expandafter\xdef\csname orcid\x\endcsname{\noexpand\href{https://orcid.org/\csname orcidauthor\x\endcsname}
			{\noexpand\orcidicon}}
}

\newcommand{\be}{\begin{equation}}
\newcommand{\ee}{\end{equation}}
\newcommand{\bea}{\begin{eqnarray}}
\newcommand{\eea}{\end{eqnarray}}

\newcommand{\beq}{\begin{equation}}
\newcommand{\eeq}{\end{equation}}

\begin{document}


\title{Gravitational Wave Spectral Shapes as a probe of Long Lived Right-handed Neutrinos, Leptogenesis and Dark Matter: \\ \it{Global versus Local $B-L$ Cosmic Strings}}

\author{Satyabrata Datta\orcidA{}}
\email{amisatyabrata703@gmail.com}
\affiliation{Institute of Theoretical Physics and Institute of Physics Frontiers
and Interdisciplinary Sciences, Nanjing Normal University}
\affiliation{Nanjing Key Laboratory of Particle Physics and Astrophysics, Nanjing 210023, China}

\author{Anish Ghoshal\orcidB{}}
\email{anish.ghoshal@fuw.edu.pl}
\affiliation{Institute of Theoretical Physics, Faculty of Physics, University of Warsaw, ul. Pasteura 5, 02-093 Warsaw, Poland}

\author{Angus Spalding \orcidC{}}
\email{angus.spalding1@gmail.com}
\affiliation{School of Physics and Astronomy, University of Southampton,
Southampton SO17 1BJ, United Kingdom}

\author{Graham White \orcidD{}}
\email{graham.white@gmail.com}
\affiliation{School of Physics and Astronomy, University of Southampton,
Southampton SO17 1BJ, United Kingdom}

\begin{abstract}
The scale of the seesaw mechanism is typically much larger than the electroweak scale. This hierarchy can be naturally explained by $U(1)_{B-L}$ symmetry, which after spontaneous symmetry breaking, simultaneously generates Majorana masses for neutrinos and produces a network of cosmic strings. Such strings generate a gravitational wave (GW) spectrum which is expected to be almost uniform in frequency unless there is a departure from the usual early radiation domination. We explore this possibility in Type I, II and III seesaw frameworks, finding that only for Type-I, long-lived right-handed neutrinos (RHN) may provide a period of early matter domination for parts of the parameter space, even if they are thermally produced. Such a period leaves distinctive imprints in the GW spectrum in the form of characteristic breaks and a knee feature, arising due to the end and start of the periods of RHN domination. These features, if detected, directly determine the right-handed neutrino mass $M$, and associated left-handed effective neutrino mass $\tilde m$ of the dominating RHN.  We find that GW detectors like LISA and ET could probe RHN masses in the range $M\in[0.1,10^{9}]$ GeV and effective neutrino masses in the $\tilde m\in[10^{-10},10^{-8}]$ eV range. We investigate the phenomenological implications of long-lived right-handed neutrinos for both local and global $U(1)_{B-L}$ strings, focusing on dark matter production and leptogenesis. We map the viable and detectable parameter space for successful baryogenesis and asymmetric dark matter production from right-handed neutrino decays. We derive analytical and semi-analytical relations correlating the characteristic gravitational-wave frequencies to the neutrino parameters $\tilde m$ and $M$, as well as to the relic abundances of dark matter and baryons. We find that the detectable parameter space reaches the boundary of hierarchical leptogenesis and encompasses a substantial portion of the near-resonant regime.

\end{abstract}
\maketitle

\tableofcontents
\section{Introduction}
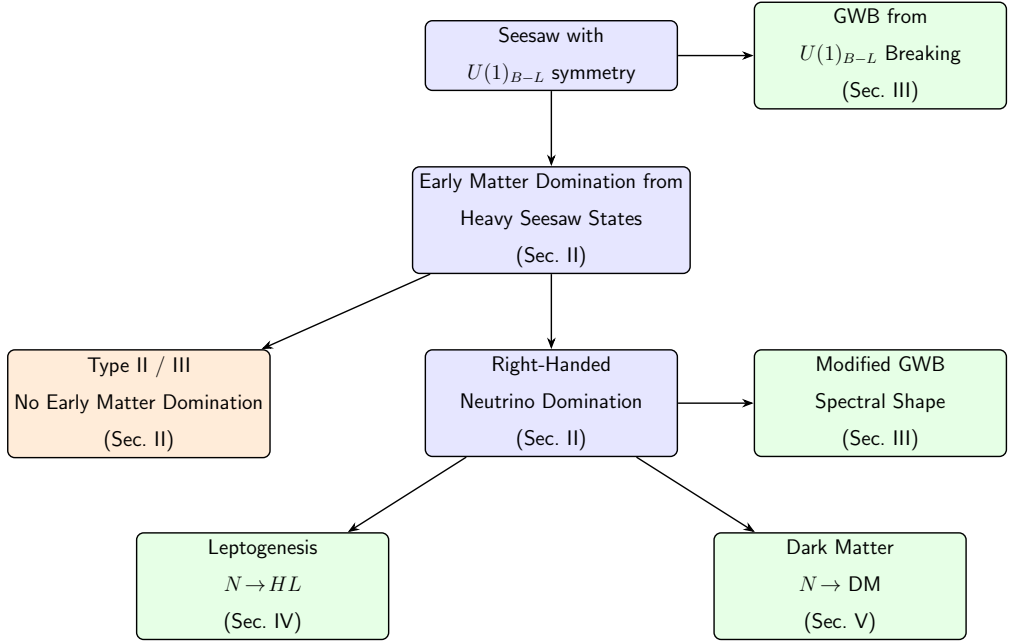
\begin{figure}[H]
\centering
\resizebox{0.8\linewidth}{!}{
\begin{tikzpicture}[
    >=Stealth,
    every node/.style={
        font=\sffamily\normalsize,
        align=center,
        rectangle,
        rounded corners,
        draw=black,
        thick,
        minimum width=5cm,
        minimum height=1.2cm
    },
    process/.style={fill=blue!10},
    result/.style={fill=green!10},
    deadend/.style={fill=orange!15},
    arrow/.style={->, thick},
    node distance=2.8cm and 2.8cm
]

\node[process] (origin) {Seesaw with \\$U(1)_{B-L}$ symmetry};
\node[result, right=1.5cm of origin] (gwb1) {GWB from\\ $U(1)_{B-L}$ Breaking\\ (Sec.~\ref{sec:cosmicstrings})};

\node[process, below=1.5cm of origin] (md) {Early Matter Domination from\\ Heavy Seesaw States\\ (Sec.~\ref{sec:domination})};

\node[deadend, below left=1.5cm and 2.8cm of md] (fail) {Type II / III\\No Early Matter Domination \\ (Sec.~\ref{sec:domination})};
\node[process, below=1.5cm of md] (rhd) {Right-Handed \\ Neutrino Domination\\ (Sec.~\ref{sec:domination})};
\node[result, right=1.5cm of rhd] (gwb2) {Modified GWB\\ Spectral Shape \\(Sec.~\ref{sec:cosmicstrings})};

\node[result, below left=1.5cm and 0.7cm of rhd] (lep) {Leptogenesis\\$N\!\to\!HL$\\ (Sec.~\ref{sec:leptogenesis})};
\node[result, below right=1.5cm and 0.7cm of rhd] (dm) {Dark Matter\\$N\!\to$ DM\\ (Sec.~\ref{sec:dark matter})};

\draw[arrow] (origin) -- (md);
\draw[arrow] (origin) -- (gwb1);
\draw[arrow] (md) -- (fail);
\draw[arrow] (md) -- (rhd);
\draw[arrow] (rhd) -- (gwb2);
\draw[arrow] (rhd) -- (lep);
\draw[arrow] (rhd) -- (dm);

\end{tikzpicture}%
}

\caption{\it The $U(1)_{B-L}$ symmetry breaking generates heavy seesaw states and an initial gravitational-wave background.
These heavy states can induce a period of early matter domination; in particular, only the Type-I seesaw, where the heavy seesaw state (the right-handed neutrino) dominates the energy density, can achieve such early matter domination. This leads to a modified gravitational-wave background (GWB) spectral shape, while the subsequent decays of the right-handed neutrinos can explain the baryon asymmetry via leptogenesis or, alternatively and simultaneously, produce dark matter.}
\label{fig: schematic}
\end{figure}

Phase transitions \cite{Mazumdar_2019, Athron:2023xlk, quiros1999finitetemperaturefieldtheory} associated with the spontaneous breaking of global or gauged $U(1)_{B-L}$ symmetries generically lead to the formation of cosmic strings \cite{Vilenkin:2000jqa}. These one-dimensional topological defects persist after formation and emit gravitational radiation through loop production and decay, sourcing a stochastic gravitational wave background (GWB) \cite{Fu:2023nrn, King:2020hyd, Ghoshal:2023sfa, Dror:2019syi}. The spectrum of this background is sensitive not only to the symmetry-breaking scale but also to the intervening cosmological history, making string-sourced GWBs a powerful probe of early Universe dynamics \cite{Cui:2017ufi, Cui:2018rwi, Gouttenoire:2019kij,Gouttenoire:2019rtn,Blasi:2020mfx,Ghoshal:2023sfa,Datta:2025yow, Roshan:2024qnv}.\\
In standard scenarios, the string network evolves during radiation domination \cite{Cui:2018rwi} and the resulting gravitational wave spectrum is almost uniform over many decades of frequency. Any departure from such a flat spectrum speaks to a surprise in our cosmic history \cite{Cui:2018rwi,Ghoshal:2023sfa,Blasi:2020wpy,Ferrer:2023uwz} with a period of early matter domination providing a particularly striking feature in the otherwise flat spectrum \cite{Cui:2017ufi,Ghoshal:2023sfa, Datta:2020bht, Samanta:2021zzk, Chianese:2024gee, Datta:2024bqp}. Specifically, the transient era of early matter domination imprints two distinct high-frequency features: a transition from a flat to a power-law spectrum and a knee arising from the superposition of modes during the transition. Together, they encode the onset and duration of the matter-dominated phase, revealing information about its physical origin. With so many decades of frequency probed by current and planned gravitational wave detectors \cite{Aasi:2014mqd,Yagi:2011wg,Punturo:2010zz,Hild:2010id,Evans:2016mbw,Sesana:2019vho}, there is a lot of opportunity for one or both observables to be detected in the foreseeable future.
A period of early matter domination can be caused by a metastable, long-lived particle \cite{Coughlan:1983ci,Starobinsky:1994bd,Dine:1995uk,Moroi:1999zb,Ghoshal:2022ruy}.
While scalar fields are often assumed to be responsible for such early matter domination, in $U(1)_{B-L}$ extended seesaw models, an alternative arises naturally: the heavy right-handed neutrinos responsible for neutrino mass generation can dominate the energy density before decaying \cite{Berbig:2023yyy}. In this paper, we investigate such scenarios in detail and show this can happen for the extended type I seesaw \cite{Minkowski:1977sc, Gell-Mann:1979vob, Yanagida:1979as, Mohapatra:1980yp}, but not type II \cite{Magg:1980ut, Wetterich:1981bx, Schechter:1980gr} or type III \cite{Foot:1988aq, Ma_1998, Ma_2002}. We further derive numerical relations linking these GW spectral features to the mass and effective neutrino mass of the right-handed neutrino responsible for the transient matter-dominated era.\\
The decays of these right-handed neutrinos are closely tied to the origin of the baryon asymmetry via leptogenesis \cite{Fukugita:1986hr}. In Leptogenesis out-of-equilibrium, CP-violating decays of right-handed neutrinos can generate a lepton asymmetry which is subsequently converted into a baryon asymmetry via electroweak sphalerons \cite{Luty:1992un, Giudice_2004, Covi_1996, Buchm_ller_2005, Sakharov:1967dj, Kolb:1979qa, Khlebnikov:1988sr}. In the presence of an intermediate matter-dominated phase, this mechanism is altered in two essential ways. First, the expansion rate is modified by early matter domination, changing the dynamics of lepton asymmetry generation. Second, the entropy injected by right-handed neutrino decays dilutes the resulting asymmetry, making the final baryon-to-entropy ratio sensitive to the duration and timing of decay. Finally, right-handed neutrino decays can also furnish a dark matter production mechanism \cite{Falkowski_2011, Barman_2022}, which can result in either symmetric or asymmetric dark matter, and we explore both possibilities. 
A schematic overview of the framework is presented in Figure \ref{fig: schematic}.

\textit{This paper is organised as follows:} In section~\ref{sec:domination}, we derive the conditions under which right-handed neutrinos dominate the energy density in both global and gauged $B-L$ scenarios, and show that such domination occurs only in the Type I seesaw. We then derive relations for the onset and duration of early matter domination. Section~\ref{sec:cosmicstrings} analyses the impact of this matter-dominated epoch on the gravitational wave spectrum from cosmic strings. Section~\ref{sec:leptogenesis} examines the consequences for baryogenesis, providing an analytic and numerical treatment of lepton asymmetry generation and the lower bound on right-handed neutrino mass for successful leptogenesis. Section~\ref{sec:dark matter} explores the implications for dark matter production from right-handed neutrino decays. We conclude in Section~\ref{sec:conclusion}.

\medskip

\section{Early Matter Domination in $U(1)_{B-L}$ Seesaw}
\label{sec:domination}
The seesaw mechanism naturally accounts for the smallness of active neutrino masses by introducing heavy states: fermion singlets (right-handed neutrinos) in type-I, scalar triplets in type-II, or fermion triplets in type-III. The masses of these heavy states are inversely proportional to the active neutrino masses \cite{Minkowski:1977sc, Gell-Mann:1979vob, Yanagida:1979as, Mohapatra:1980yp, Magg:1980ut, Wetterich:1981bx, Schechter:1980gr, Foot:1988aq, Ma_1998, Ma_2002},
\begin{equation}
    m_{\nu}\propto\frac{v_H^2}{M}
\end{equation}
where $M$ is the mass of the heavy seesaw state and $v_H=246$ GeV is the Higgs vacuum expectation value (vev). A large $M$ naturally explains the smallness of active neutrino masses.  A $U(1)_{B-L}$ symmetry provides a natural explanation for why $M$ is so large: the heavy seesaw masses are tied to the scale at which $B-L$ is spontaneously broken, $v_{B-L}$, so if the scale of symmetry breaking is large, then so is the mass of the heavy seesaw state. Moreover, $U(1)_{B-L}$ is not always ad hoc, but appears in many Grand Unified Theory symmetry-breaking chains, giving further motivation for embedding the seesaw mechanism within this framework \cite{Gell-Mann:1979vob,Buchmuller:2005eh,Dror:2019syi,Dunsky:2021tih}. The particle content for each type of seesaw is shown in Table \ref{tab:BL-table}.
\begin{table}[h!]
\centering
\begin{tabular}{c|c c c c}
\hline
Field & $SU(3)_c\ $ & $SU(2)_L\ $ & $U(1)_Y\ $ & $U(1)_{B-L}$ \\
\hline
$q^i_L$     & $\mathbf{3}$ & $\mathbf{2}$ & $+1/6$ & $+1/3$ \\
$u^i_R$     & $\mathbf{3}$ & $\mathbf{1}$ & $+2/3$ & $+1/3$ \\
$d^i_R$     & $\mathbf{3}$ & $\mathbf{1}$ & $-1/3$ & $+1/3$ \\
$\ell^i_L$  & $\mathbf{1}$ & $\mathbf{2}$ & $-1/2$ & $-1$ \\
$e^i_R$     & $\mathbf{1}$ & $\mathbf{1}$ & $-1$   & $-1$ \\
$H$         & $\mathbf{1}$ & $\mathbf{2}$ & $-1/2$ & $0$ \\
\hline
$\Phi$      & $\mathbf{1}$ & $\mathbf{1}$ & $0$    & $+2$ \\
$N^i_R$ (Type I)     & $\mathbf{1}$ & $\mathbf{1}$ & $0$   & $-1$ \\
$\Delta$ (Type II)   & $\mathbf{1}$ & $\mathbf{3}$ & $+1$  & $-2$ \\
$\Sigma$ (Type III)  & $\mathbf{1}$ & $\mathbf{3}$ & $0$   & $-1$ \\
$Z'$ (local)  & $\mathbf{1}$ & $\mathbf{1}$ & $0$   & $0$ \\
\hline
\end{tabular}
\caption{\it Particle content of the $B-L$ extended Seesaw Model. The first block lists the Standard Model particles. The second block shows the additional scalar $\Phi$ required to break $U(1)_{B-L}$ and the heavy states associated with type I, II, and III seesaw mechanisms. If the symmetry is local, there is also a corresponding $Z'$ gauge boson.}
\label{tab:BL-table}
\end{table}

\subsection{The Condition for Early Matter Domination}
In this section, we derive the analytic conditions for right-handed neutrino early matter domination. We first present an approximate derivation, following the standard approach in the literature. While convenient, this estimate is inaccurate by roughly half an order of magnitude. We therefore provide a more accurate derivation, which can still be carried out analytically.

\subsubsection{Derivation I: Canonical derivation}
We consider a particle species \( N \) of mass \( M \) that decouples from the thermal bath while still relativistic. 
At high temperatures \( T \gg M \), both the particle and radiation energy densities scale identically with temperature, so their ratio reduces to a ratio of degrees of freedom.
\begin{equation}
\rho_N = \frac{\pi^2}{30} \, g \, T^4,\quad \rho_R=\frac{\pi^2}{30} \, g_* \, T^4,\quad \frac{\rho_N}{\rho_R}(T \gg M) = \frac{g}{g_*}
\end{equation}
where \( g,\ g_* \) denote the species and radiation degrees of freedom, respectively. As the Universe cools and \( T \lesssim M \), the species \( N \) becomes non-relativistic and its energy density redshifts as matter, \( \rho_N \propto a^{-3} \), while radiation continues to redshift as \( \rho_R \propto a^{-4} \). Consequently, the energy density ratio increases as the temperature decreases,
\begin{equation}
\frac{\rho_N}{\rho_R}(T \ll M) \approx \frac{g}{g_*} \ \frac{M}{T}.
\end{equation}
Matter domination occurs when \( \rho_N = \rho_R \), which defines the domination temperature:
\begin{equation}
T_{\text{dom}} \approx \frac{g}{g_*} \ M \ .
\end{equation}
As a concrete example, the Type-I seesaw introduces Standard Model singlets \( N_i \) with degrees of freedom \( g = \tfrac{7}{8} \times 2 \). This gives \cite{Giudice_1999}
\begin{equation}
    T_{\text{dom}} \simeq 0.016 \, M\ .
\end{equation}
This illustrates how the onset of early matter domination in seesaw scenarios is directly controlled by the heavy state mass.

\subsubsection{Derivation II: Comoving Abundance Method}
A more accurate approach is to work directly with comoving abundances. This avoids any assumptions about the point at which the species changes from radiation-like to matter-like behaviour, since the analysis is performed at very high and very low temperatures relative to the mass. Consider a particle species $N$ of mass $M$ that decouples while still relativistic. The number and entropy densities of a relativistic species are both proportional to temperature cubed, giving a temperature-independent yield that has frozen out.
\begin{equation}
n(T) = \frac{\zeta(3)}{\pi^2}\, g \, T^3 ,\quad s(T) = \frac{2\pi^2}{45}\, g_*(T)\,T^3, \quad Y \;\equiv\; \frac{n}{s} =0.0026 g\ .
\end{equation}
Once $T \ll M$, the species behaves as non-relativistic matter and its energy density is well approximated by $\rho_N = M\,Y\,s(T)$. In this regime, the average particle energy, averaging over momentum space, is $\langle E \rangle \simeq M + \mathcal{O}(T)$, so thermal corrections are negligible compared to the mass. Equating the right-handed neutrino and radiation energy densities $\rho_N = \rho_R$, sets the temperature for early matter domination
\begin{equation}
T_{\rm dom} = \frac{4}{3}\, M Y \simeq 0.37\,\frac{g}{g_*}\,M\ .
\end{equation}
This result differs by a factor of three compared to the first derivation. For right-handed neutrinos, this yields
\begin{equation}
    Y_N^i=3.9\times 10^{-3},\qquad T_{\rm dom}^N=0.52\% M,
    \label{eq: RHN Tdom}
\end{equation}
where the superscript $i$ denotes the initial value. In what follows, we adopt Derivation~II. For successful early matter domination by such particles, the following three conditions must be satisfied:
\begin{enumerate}
    \item The particle must be thermally produced relativistically
    \begin{equation}
        \Gamma_{\text{prod}} > H(T \gg M)\ .
        \label{eq:Cond I}
    \end{equation}
    where $H$ is the Hubble parameter.
    \item The particle must freeze-out before becoming non-relativistic:
    \begin{equation}
        \Gamma_{\text{ann}}(T \approx M) \sim \Gamma_{\text{prod}}(T \approx M) < H(T\approx M).
        \label{eq:Cond II}
    \end{equation}
    This prevents the abundance from being Boltzmann suppressed once $T < M$, allowing the species to retain a relic density large enough to eventually dominate.
    \item The particle must not have decayed by the time of early  matter domination:
    \begin{equation}
        \Gamma_{\text{decay}} < H(T_{\text{dom}}).
    \end{equation}
\end{enumerate}
We now investigate these three conditions for the $U(1)_{B-L}$ extended seesaw frameworks. If instead we consider non-thermal production channels, such as inflaton decay \cite{Asaka_1999,Hahn_Woernle_2009,Ghoshal:2022fud}, curvaton decay \cite{Fong:2023egk}, generic and modulated sneutrino decay \cite{Mazumdar:2003va,Allahverdi:2002gz,Mazumdar:2003bs,Afzal:2024hwj}, Q-ball decay \cite{Fujii:2001sn}, from phase transition bubbles \cite{Dasgupta:2022isg,Cataldi:2024pgt}, preheating \cite{Cui:2024vws} and reheating \cite{Ghoshal:2022kqp}, or primordial black holes evaporations \cite{Datta:2020bht, das2021lowscaleleptogenesisdark, Perez_Gonzalez_2021, Barman_2022, Bernal_2022, Ghoshal:2023sfa}, only conditions (ii) and (iii) need to be imposed. To remain general, we shall leave the initial abundance $Y_i$ as a free parameter throughout our analysis.

\subsection{Condition I and II: Thermal Production and Relativistic Freeze-out}
\subsubsection{Type I: Global $U(1)_{B-L}$ case}
The type~I seesaw mechanism \cite{Minkowski:1977sc, Gell-Mann:1979vob, Yanagida:1979as} is among the most economical and widely studied explanations for the origin of light neutrino masses. In this framework, heavy right-handed neutrinos $N_i$ are introduced, which couple to the Standard Model lepton doublets $L_\alpha$ and the Higgs doublet $H$ through Yukawa interactions. After electroweak symmetry breaking, these interactions generate a Dirac mass term for the neutrinos. If the right-handed neutrinos also acquire large Majorana masses, the light neutrinos obtain naturally small masses via the seesaw mechanism. Since the Majorana mass term explicitly violates $B-L$ by two units, the global $U(1)_{B-L}$ symmetry requires the introduction of an additional complex scalar field $\Phi$, carrying non-zero $B-L$ charge \cite{Fu:2023nrn}. When $\Phi$ acquires a vacuum expectation value, the $B-L$ symmetry is broken, the right-handed neutrinos obtain Majorana masses, and the seesaw mechanism is realized. The Lagrangian is augmented by Yukawa, Majoron and Higgs portal interactions, as well as a quartic potential for the symmetry-breaking scalar
\begin{equation}
\begin{aligned}
    \mathcal{L}_{\rm Type\ I} \supset& -y_{\alpha i}L_{\alpha}\tilde{H}N_i
    -\tilde y_i\Phi\,\overline{N_i^C}N_i \\
    &-m_{\phi}|\Phi|^2
    -\lambda_\phi |H|^2|\Phi|^2
    -\lambda_4|\Phi|^4
\end{aligned}
\label{eq: TypeI Lagrangian}
\end{equation}
where $\alpha$ labels the lepton flavour, $i$ labels the right-handed neutrino species, $L_\alpha$ denotes the SM lepton doublet, $H$ the Higgs doublet, and $N_i$ the right-handed neutrino singlet. After $\Phi$ acquires a vev, $v_{B-L}$, the production of right-handed neutrinos proceeds via mediation of the real scalar excitation $\mathfrak{Re} (\Phi) = \phi$:
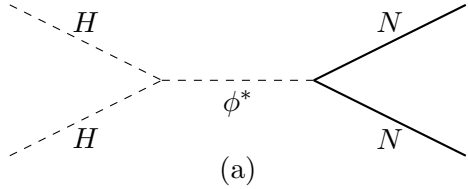
\begin{figure}[h!]
\centering
\begin{tikzpicture}
  \coordinate (H1) at (-3,1);
  \coordinate (H2) at (-3,-1);

  \coordinate (v1) at (-1,0);

  \coordinate (v2) at (1,0);

  \coordinate (N1) at (3,1);
  \coordinate (N2) at (3,-1);

  \draw[dashed] (H1) -- (v1) node[midway, above] {$H$};
  \draw[dashed] (H2) -- (v1) node[midway, below] {$H$};
  \draw[dashed] (v1) -- (v2) node[midway, below] {$\phi^*$};
  \draw[thick] (v2) -- (N1) node[midway, above] {$N$};
  \draw[thick] (v2) -- (N2) node[midway, below] {$N$};

  \node at (0,-1.2) {(a)};
\end{tikzpicture}
\caption{\it Thermal production of right-handed neutrinos in global $U(1)_{B-L}$ model. The s-channel production of right-handed neutrinos proceeds through the exchange of the real scalar antiparticle $\phi^*$.}
\label{fig:HHtoNN}
\end{figure}
For off-shell exchange where the centre of mass energy is much less than the mass of the scalar field, the production rate increases more slowly with temperature than the expansion rate
\begin{equation}
\Gamma_{\text{prod}} \sim \left( \frac{\lambda_\phi\ v_{B-L}\ \tilde y_i}{m_\phi^2} \right)^2 T, \quad \Rightarrow \quad \frac{\Gamma_{\rm prod}}{H} \sim \left( \frac{\lambda_\phi\ v_{B-L}\ \tilde y_i }{m_\phi^2} \right)^2 \ \frac{M_{\text{Pl}}}{T} \ .
\end{equation}
Therefore, we cannot have production at an ultra-relativistic regime and relativistic freeze-out. In the opposite regime, the scalar propagator is highly suppressed and the interaction rate falls more rapidly,
\begin{equation}
\frac{\Gamma_{\rm prod}}{H} \propto \frac{1}{T^5}
\end{equation}
Since this ratio decreases with temperature, we conclude that production is inefficient for the global case. We therefore conclude that thermal production (Eq. \ref{eq:Cond I}) and relativistic freeze-out (Eq. \ref{eq:Cond II})
, in the global $U(1)_{B-L}$ case cannot occur by thermal production; however, it can occur with non-thermal production.
\subsubsection{Type I: Local $U(1)_{B-L}$ case}
Insisting on a local $U(1)_{B-L}$ necessitates the existence of a gauge boson $Z'$ which couples to both right-handed neutrinos and standard model fermions with a coupling $g_{B-L}$. The dominant process of right-handed neutrino production is SM fermion annihilation via $Z'$ exchange:
\begin{figure}[h!]
\centering
\begin{tikzpicture}[decoration={markings, mark=at position 0.5 with {\arrow{>}}}]

  \coordinate (f) at (-3,1);
  \coordinate (fbar) at (-3,-1);

  \coordinate (v1) at (-1,0);

  \coordinate (v2) at (1,0);

  \coordinate (N) at (3,1);
  \coordinate (Nbar) at (3,-1);

  \draw[thick,postaction={decorate}] (f) -- (v1) node[midway, above] {$f$};
  \draw[thick,postaction={decorate}] (fbar) -- (v1) node[midway, below] {$\bar f$};

  \draw[decorate,decoration=snake] (v1) -- (v2) node[midway, above] {$Z'$};

  \draw[thick] (v2) -- (N) node[midway, above] {$N$};
  \draw[thick] (v2) -- (Nbar) node[midway, below] {$\bar N$};

  \node at (0,-1.5) {(a)};
\end{tikzpicture}
\caption{\it Thermal Production of right-handed neutrinos via a $Z'$ mediator.}
\label{fig:ffbarToNN}
\end{figure}
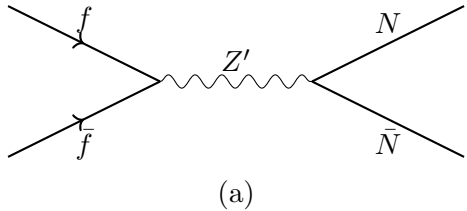
We again assume all external particles are relativistic ($T \gg m_f, M$) and the mediator is off-shell. The production rate then increases faster with temperature than the expansion rate 
\begin{equation}
\frac{\Gamma_{\text{prod}}}{H} \sim \frac{T^3 M_{\text{Pl}}}{v_{B-L}^4},
\end{equation}
so production and freeze-out are easily achieved for sufficiently large temperatures. For the opposite regime, we cannot satisfy these conditions. We therefore have that for a large $M_{Z'}$ we can satisfy conditions I and II, Eqs \ref{eq:Cond I} \ref{eq:Cond II}.
\subsubsection{Type II and III seesaw}
The Type-I seesaw is the simplest way to generate light neutrino masses via the seesaw mechanism, but it is not the only one. Neutrino masses can also be generated through the Type-II seesaw \cite{Magg:1980ut, Wetterich:1981bx, Schechter:1980gr}, which introduces an electroweak scalar triplet, and the Type-III seesaw \cite{Foot:1988aq, Ma_1998, Ma_2002}, which introduces fermionic triplets. We now turn to the thermal history of these triplet states to examine the conditions under which early matter domination could occur. Both the type II and III seesaw mechanisms include a BSM particle which is electroweakly charged. The dominant contribution arises from $s$-channel scattering mediated by the electroweak $Z$ boson:
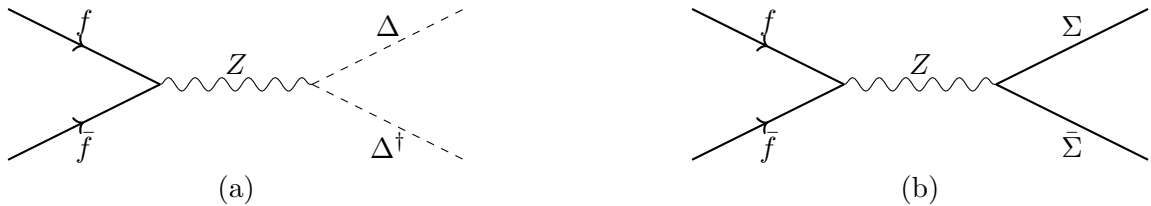
\begin{figure}[h!]
\centering
\begin{tikzpicture}[decoration={markings, mark=at position 0.5 with {\arrow{>}}}]

  \coordinate (f1) at (-3,1);
  \coordinate (fbar1) at (-3,-1);
  \coordinate (v1) at (-1,0);
  \coordinate (v2) at (1,0);
  \coordinate (d1) at (3,1);
  \coordinate (d2) at (3,-1);

  \draw[thick,postaction={decorate}] (f1) -- (v1) node[midway, above] {$f$};
  \draw[thick,postaction={decorate}] (fbar1) -- (v1) node[midway, below] {$\bar f$};
  \draw[decorate,decoration=snake] (v1) -- (v2) node[midway, above] {$Z$};
  \draw[dashed] (v2) -- (d1) node[midway, above] {$\Delta$};
  \draw[dashed] (v2) -- (d2) node[midway, below] {$\Delta^\dagger$};
  \node at (0,-1.4) {(a)};

  \begin{scope}[xshift=9cm]
    \coordinate (f2) at (-3,1);
    \coordinate (fbar2) at (-3,-1);
    \coordinate (v3) at (-1,0);
    \coordinate (v4) at (1,0);
    \coordinate (s1) at (3,1);
    \coordinate (s2) at (3,-1);

    \draw[thick,postaction={decorate}] (f2) -- (v3) node[midway, above] {$f$};
    \draw[thick,postaction={decorate}] (fbar2) -- (v3) node[midway, below] {$\bar f$};
    \draw[decorate,decoration=snake] (v3) -- (v4) node[midway, above] {$Z$};
    \draw[thick] (v4) -- (s1) node[midway, above] {$\Sigma$};
    \draw[thick] (v4) -- (s2) node[midway, below] {$\bar \Sigma$};
    \node at (0,-1.4) {(b)};
  \end{scope}

\end{tikzpicture}
\caption{\it Dominant $s$–channel processes mediated by the electroweak $Z$ boson: 
(a) production of scalar triplets $\Delta\Delta^\dagger$ in the type~II seesaw, 
(b) production of fermion triplets $\Sigma\bar\Sigma$ in the type~III seesaw. }
\label{fig:ffbarToSeesaw}
\end{figure}
The production rate in this case grows more slowly with temperature than the expansion rate
\begin{equation}
\frac{\Gamma_{\text{prod}}}{H} \sim \frac{g_Z^4 T}{T^2 / M_{\text{Pl}}} = g_Z^4 \ \frac{M_{\text{Pl}}}{T}.
\end{equation}
With $g_Z$ being large for many species in the Standard Model \cite{ParticleDataGroup:2016lqr, ParticleDataGroup:2020ssz} and summing over the possible diagrams, we cannot have freeze out for temperatures below $10^{18}\rm GeV$ and therefore right-handed neutrino masses below the Planck scale, and as such we can never satisfy condition II (Eq. \ref{eq:Cond II}). We conclude we cannot have early matter domination from the seesaw heavy state in type II and III for thermal or non-thermal production. This is true for any model with Standard Model interactions; BSM interactions, for instance, mediated by $Z^{\prime}$ in the $B-L$ extension, can only exacerbate the problem.

\subsection{Condition III: Late decays}
\noindent In type I see-saw right-handed neutrino decay rate is proportional to the mass and the squared Yukawa couplings,
\begin{equation}
    \Gamma_i=\frac{(y^\dagger y)_{ii}M_i}{8\pi}\ .
    \label{eq: Decay rate}
\end{equation}
Here $i$ labels the heavy right-handed neutrino mass eigenstate, running over $i=1,2,3$, and we have taken the masses of the Higgs and Leptons to be negligible compared to the mass of the right-handed neutrinos. This decay rate must be smaller than the Hubble rate at the matter-domination temperature
\begin{equation}
    H(T_{\rm dom})=\sqrt{\frac{8\pi}{3M_{\rm Pl}^2}\,(\rho_N+\rho_R)}=\sqrt{\frac{8\pi}{3M_{\rm Pl}^2}\,2\rho_R(T_{\rm dom})}
\end{equation}
where $M_{\rm Pl}$ is the Planck mass. We adopt the Casas-Ibarra parameterisation of the Yukawa matrix \cite{Casas_2001}, and it emerges that this is suppressed by the masses of the active neutrinos.
\begin{equation}
    y=\frac{1}{v_H}U\sqrt{m}R^T\sqrt{M},\quad (y^\dagger y)_{ii} = \frac{M_i}{v_H^2} \sum_{j=1}^3 m_j |R_{ij}|^2=\frac{M_i\tilde{m_i}}{v_H^2} 
\end{equation}
where $m$ and $M$ are the diagonal matrices of light and heavy neutrino masses, respectively, and $\tilde{m_i}$ denotes the effective neutrino mass. The matrix $U$ is the standard PMNS matrix, and $R$ is a complex orthogonal matrix. This suppression of the decay rate gives us a fighting chance of sufficiently delaying decays to achieve early matter domination. With this parameterisation, the condition for right-handed neutrino early matter domination reduces to a simple requirement on the effective neutrino mass.
\begin{equation}
    \tilde{m_i}=\sum_{j=1}^3 m_j |R_{ij}|^2 < 2.89 \times 10^{-17}\, \mathrm{GeV}
\end{equation}
Using the orthogonality condition $\sum_{j}|R_{ij}|^2=1$
one can easily show $\text{Min}(\tilde{m_i})=\text{Min}(m_{i})$. Thus, the requirement for right-handed neutrino early matter domination reduces to a bound on the lightest active neutrino mass,
\begin{equation}
    m_{\text{lightest}}<2.89\times 10^{-8}\ \rm eV.
    \label{eq: mtilde bound}
\end{equation}
This result is independent of active neutrino mass ordering. An analytical estimate has been presented here, while the corresponding numerical results confirming this behaviour are shown later in Fig.~\ref{fig:MD_scans}.

\subsection{Decay Temperature}
\noindent To determine the duration of the right-handed neutrino-dominated era, we must identify when the species decays and transfers its energy back into radiation. This occurs at the decay temperature $T_{\rm dec}$, defined by $\Gamma = H(T)$, where we assume the decay to be instantaneous. We recall that the ratio of right-handed neutrino to radiation energy densities evolves as a ratio of temperatures
\begin{equation}
    \frac{\rho_N(T)}{\rho_R(T)} = \frac{4YM}{3T}=\frac{T_{\rm dom}}{T},
    \label{eq: rho to T}
\end{equation}
we can rewrite the total energy density and therefore the Hubble rate in terms of this ratio,
\begin{equation}
    H(T) = 1.66\,\sqrt{g_*}\,\frac{T^2}{M_{\rm Pl}}
           \sqrt{1 + \frac{T_{\rm dom}}{T}}.
\end{equation}
Equating this to the decay rate, yields
\begin{equation}
    T_{\rm dec}^4 + T_{\rm dom}\,T_{\rm dec}^3 -
    \left[ \frac{\Gamma\,M_{\rm Pl}}
                {1.66 \sqrt{g_*}} \right]^2 = 0.
\end{equation}
Since the exact analytic solution is unwieldy, we instead consider the matter-dominated decay limit, $T_{\rm dom} \gg T_{\rm dec}$, in which the expression simplifies to
\begin{equation}
    T_{\rm dec} \simeq \left(\frac{C^2}{T_{\rm dom}}\right)^{1/3},
    \qquad
    C \equiv \frac{\Gamma M_{\rm Pl}}{1.66 \sqrt{g_*}},
\end{equation}
For the decay rate of a Majorana right-handed neutrino, we have
\begin{equation}
    T_{\rm dec} \simeq 2.55\times 10^{2}\;
    \left(\frac{\tilde m}{\mathrm{eV}}\right)^{\!2/3}
    \left(\frac{M}{\mathrm{GeV}}\right)\ \mathrm{GeV}.
    \label{eq: Tdec}
\end{equation}
The condition on the effective neutrino mass ensures that decays occur after the right-handed neutrino dominates the energy budget. The duration is then written in terms of the number of e-folds of early matter domination, $N_e$, which depends only on the effective neutrino mass,
\begin{equation}
    N_e=\ln\left(\frac{T_{\rm start}}{T_{\rm end}}\right)\simeq\ln\left(\frac{T_{\rm dom}}{T_{\rm dec}}\right)\simeq \ln \!\left( 2.04 \times 10^{-5}\, \left(\frac{\tilde{m}}{\rm eV}\right)^{-\tfrac{2}{3}} \right),
    \label{eq: Nefolds}
\end{equation}
whereas the onset of early matter domination is uniquely determined by the right-handed neutrino mass $M$. The numerics of this is also shown later in Fig.~\ref{fig:MD_scans}.

\subsection{Duration of Early Matter Domination}
\begin{figure}[H]
\centering
\includegraphics[width=0.52\linewidth]{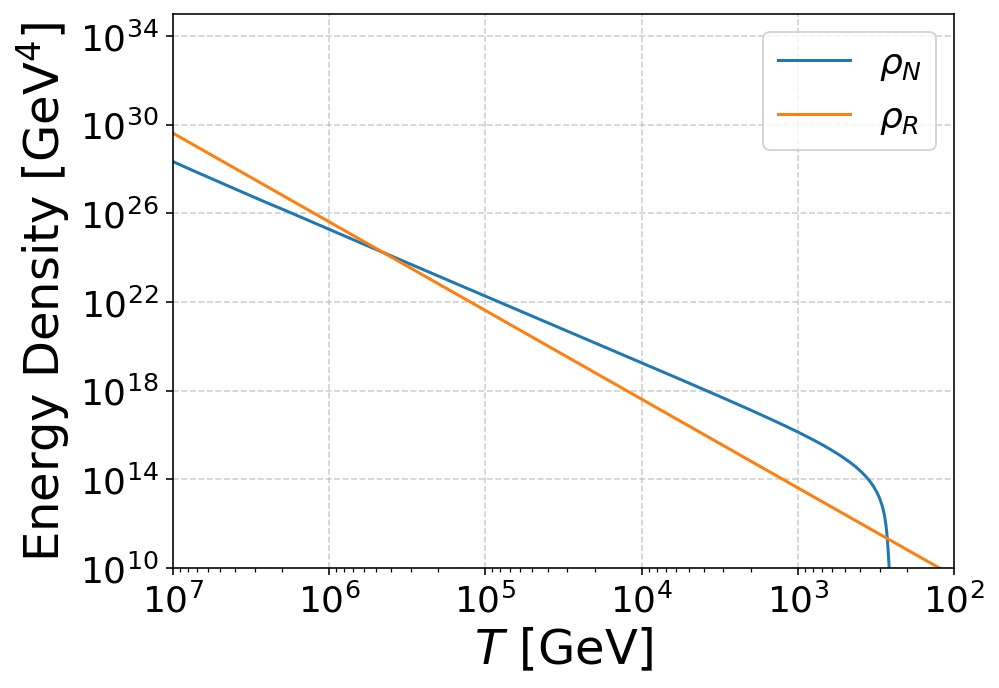} 
\includegraphics[width=0.47\linewidth]{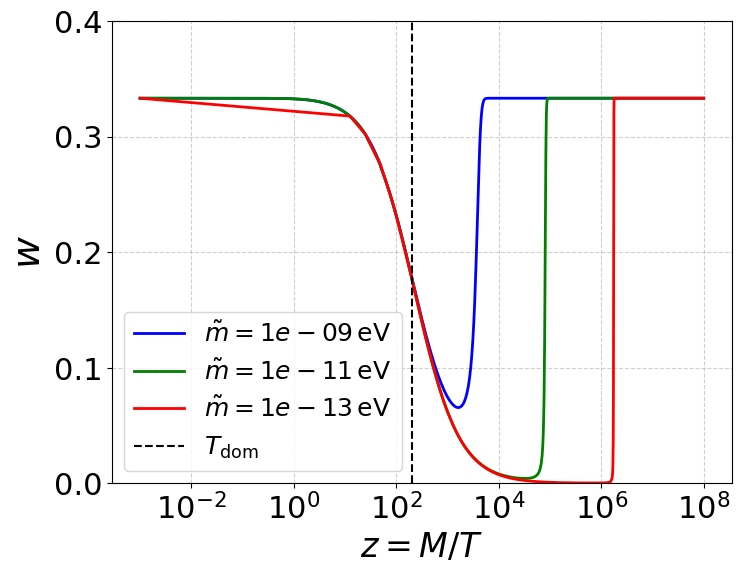} 
\caption{\it \textbf{Left Panel:} Energy density evolution for $\tilde m=10^{-10}\ \text{eV},\ M=10^9\ \text{GeV}$ showing how the right-handed neutrino comes to dominate the energy budget of the universe. \textbf{Right Panel:} effective equation of state $w(z)$ for a few effective neutrino mass values. $\rho_N$ becomes the dominant component of the energy budget at the vertical dashed line.}
    \label{fig:MD benchmark}
\end{figure}

To determine the duration of the early matter–dominated era, we solve the coupled Boltzmann system for the entropy–normalised abundances $Y\equiv n/s$. For a right-handed neutrino undergoing the two–body decay $N\to 2R$, where R denotes relativistic radiation particles, the corresponding equations read
\begin{align}
    \frac{dY_N}{dz} &= -D(z)\,\Big(Y_N - Y_N^{\rm eq}(z)\Big), \label{eq:NNfull}\\
    \frac{dY_R}{dz} &= 2\,D(z)\,\Big(Y_N - Y_N^{\rm eq}(z)\Big), \label{eq:NRfull}
\end{align}
where $z=M/T$ is the standard dimensionless variable parametrising the evolution. $Y_N^{\rm eq}$ and $D$ are the normalised equilibrium abundance and the parameter describing the decay, respectively
\begin{equation}
    Y_N^{\rm eq}(z)=\frac{45\,g_N}{4\pi^4 g_*}\,z^2 K_2(z)\ \quad  D(z) = \frac{\Gamma}{zH(z)}\ ,
    \label{eq: D}
\end{equation}
where $K_2$ is the modified Bessel function of the second kind and with initial conditions $Y_N(z_i)=Y_N^i$ and $Y_R(z_i)=Y_R^{\rm eq}$, we solve the system numerically. As a benchmark, Fig.~\ref{fig:MD benchmark} shows the resulting evolution of the right-handed–neutrino and radiation energy densities for $\tilde m=10^{-10}\ \text{eV}$ and $M=10^9\ \text{GeV}$, together with the effective equation of state for various effective neutrino mass values illustrating the dynamics of the intermediate matter-dominated period.  

\begin{figure}[H]
    \centering
    \includegraphics[width=0.49\linewidth]{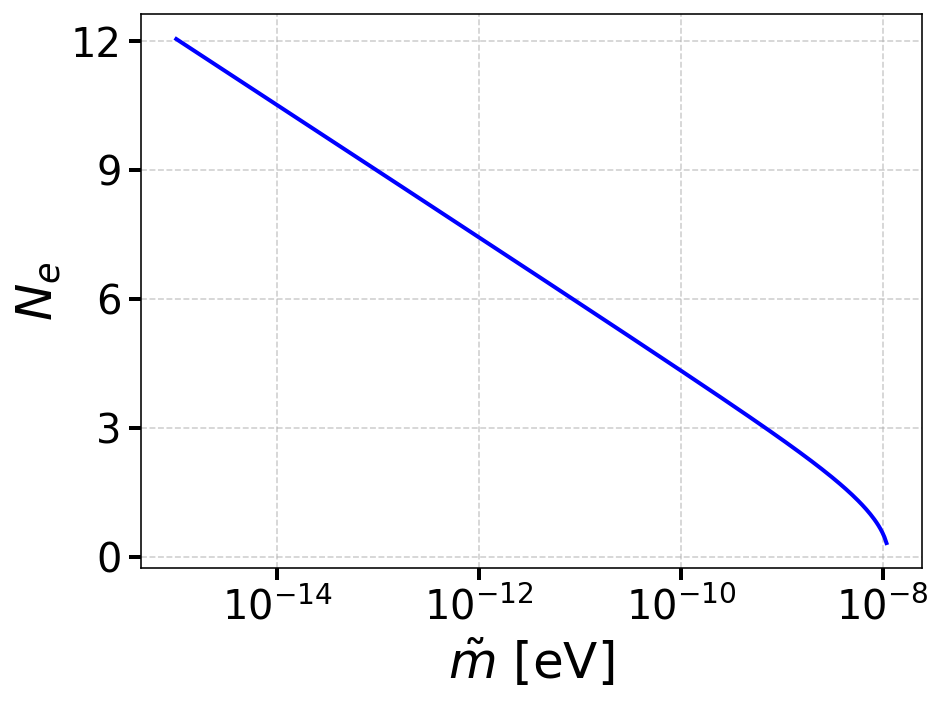}
    \includegraphics[width=0.49\linewidth]{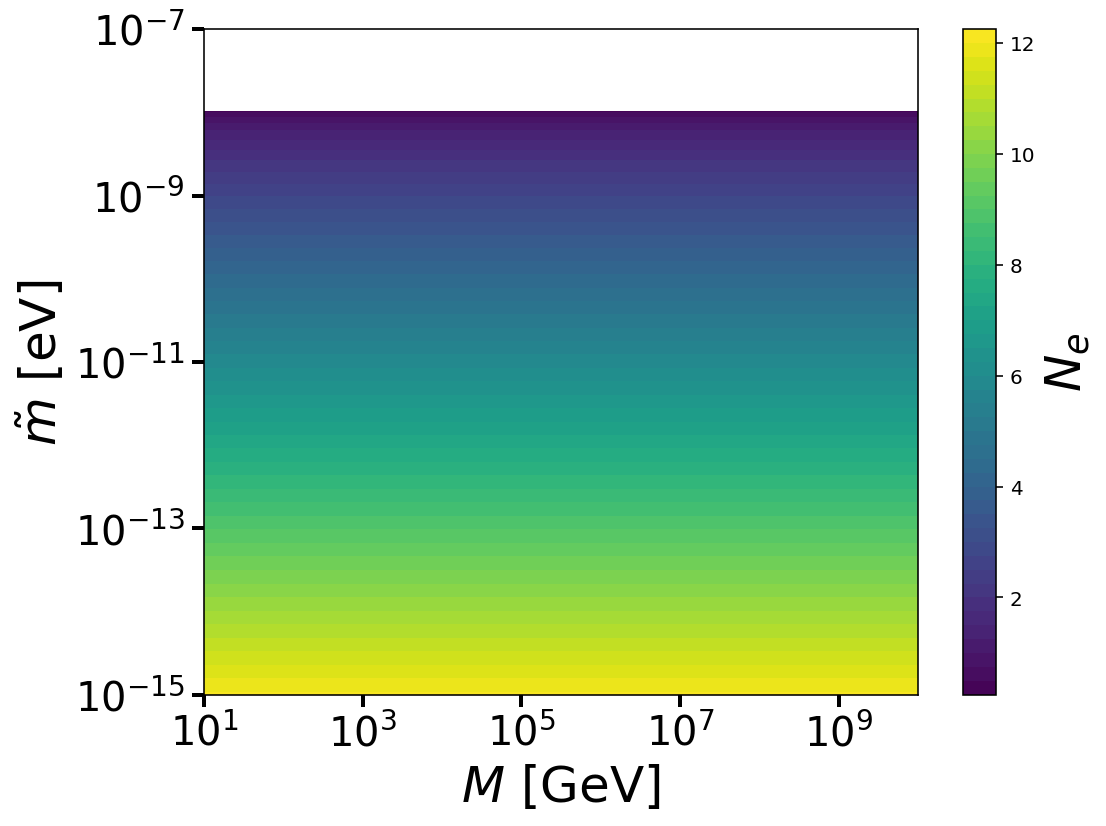}
    \caption{\it Duration of early matter domination in terms of number of e-folds of right-handed neutrino domination, $N_e$. 
    \textbf{Left:} Dependence on the effective neutrino mass. A smaller effective mass delays right-handed neutrino decays, extending the period of early matter domination.
    \textbf{Right:} dependence on both $M$ and $\tilde m$ showing early matter domination duration is independent of right-handed neutrino mass.}
    \label{fig:MD_scans}
\end{figure}

From the numerical solution, we find that the proportionality constant in the analytic estimate of $T_{\rm dom}$, Eq. \ref{eq: RHN Tdom}, was overestimated, with the numerical result smaller by a factor of about $1.2$
\begin{equation}
    T_{\rm dom}=0.45\%\, M\ .
    \label{eq: Tdom}
\end{equation}
This discrepancy arises because the analytic treatment assumes no decays occur before $\Gamma=H$, while in reality, decays begin earlier; this means it takes longer for the right-handed neutrino to dominate the energy budget. We performed a parameter scan over $(M,\tilde m)$ to extract the duration of early matter domination, $N_e$. The result is shown in Fig.~\ref{fig:MD_scans}, both as a function of $\tilde m$ and on the full $(\tilde m, \rm M)$ plane, showing the duration is essentially independent of $M$.
\noindent These results show that the onset of early matter domination is uniquely determined by the mass of the right-handed neutrino, whilst the duration is determined by the effective neutrino mass. We also found our analytic bound of effective neutrino mass for right-handed neutrino, Eq. \ref{eq: mtilde bound}, is slightly overestimated also.
\begin{equation}
    \tilde m<1.1\times 10^{-8}\rm eV
    \label{eq: numeric mtilde}
\end{equation}
$N_e$ is well described by a best–fit relation
\begin{equation}
    N_e(\tilde{m}) \;\approx\; 0.68 \, \ln\!\left(\frac{5.22 \times 10^{-8}\  \rm eV}{\tilde{m}} - 1\right)\ .
    \label{eq: MD duration}
\end{equation}
This result closely matches the analytical estimate of Eq.~\ref{eq: Nefolds} in the small-$\tilde m$ limit, exhibiting the same $\tilde m$-dependence but differing by only a factor of $\mathcal{O}(2)$ in the logarithmic prefactor. Combining this relation with the equation for $T_{\rm dom}$, Eq. \ref{eq: Tdom}, and the equation above, we obtain a formula for the end of early matter domination \footnote{Unless otherwise specified, to avoid cumbersome notation the reader should assume any unspecified units are in GeV.}. 
\begin{equation}
T_{\rm end}(M, \tilde{m}) \;\approx\; 
4.5 \times 10^{-3} \, M \,
\left( \frac{5.22 \times 10^{-8}\ \rm eV}{\tilde{m}} - 1 \right)^{-0.68}
\label{eq: Tend}
\end{equation}
In the small-$\tilde m$ limit, this reduces to the usual expression for $T_{\rm dec}$, Eq. \ref{eq: Tdec}, with the same scaling in $M$ and $\tilde m$ but differing by a factor of roughly $1.5$ in the prefactor. This shift reflects the fact that decays commence before the condition $\Gamma=H$ is exactly satisfied, leading to an earlier end of early matter domination than predicted analytically. Together, these relations provide a one–to–one mapping between the neutrino mass parameters and the thermal history of early matter domination. Crucially, we will find that the measurable quantities of the gravitational wave background spectral shape will be determined by the beginning and end temperatures of early matter domination, which we can then translate into a detectable region of the $(M,\tilde m)$ parameter space using the relations above.

\medskip

\section{Cosmic strings Gravitational Waves as a cosmic witness of long-lived neutrinos}\label{sec:cosmicstrings}

In this section, we discuss the gravitational wave footprints that long-lived neutrinos will leave during a period of early matter domination if there is a network of cosmic strings. This, of course, means taking a mild digression to discuss the theory behind gravitational wave signals from cosmic string networks, in both cases where the corresponding symmetry is global or local. After introducing the machinery of gravitational waves from strings in standard and non-standard cosmic histories, we will look at the detectability of long-lived neutrinos later in this section.

\subsection{Gravitational wave spectrum from cosmic strings in standard and non-standard cosmic histories} \label{subCS}

The spontaneous breaking of a $U(1)$ symmetry, for instance,  $U(1)_{\rm B-L}$ whether gauged or global, inevitably leads to the formation of a network of cosmic strings, a kind of topological defect in the early universe \cite{Kibble:1976sj, Vilenkin:2000jqa}. As the universe evolves, long strings undergo intercommutation, producing closed string loops that subsequently oscillate and radiate energy \cite{Vachaspati:1984gt, Allen:1991bk,Gouttenoire:2019kij, Auclair:2019wcv, Gouttenoire:2022gwi, Simakachorn:2022yjy}, predominantly via gravitational waves (gauged case) or Goldstone bosons (global case)\footnote{A persistent discrepancy remains between the precise predictions obtained from field‑theoretic analyses and those from lattice‑based string simulations; see Refs. \cite{Vincent:1997cx,Matsunami:2019fss,Hindmarsh:2021mnl,Blanco-Pillado:2023sap} for detailed discussions.}. The properties of the long string network are encapsulated by a correlation length $L=\sqrt{\mu/\rho_\infty}$, defined through the energy density $\rho_\infty$ in long strings, where $\mu$ is the string tension related to the symmetry breaking scale as \cite{Bogomolny:1975de},
\begin{equation}
\mu = 2\pi n \, v_{\rm B-L}^2 \times \begin{cases}
1 ~ ~ ~ & {\rm for ~ local ~ strings},\\
\log(v_{\rm B-L} t)  ~ ~ ~ & {\rm for ~ global ~ strings}.
\end{cases}
\label{string_tension}
\end{equation} 
Here $v_{\rm B-L}$ represents the vev of the scalar field which breaks the $U(1)_{\rm B-L}$ leading to cosmic string network formation, and $n$ is the winding number. Note, in the case of global strings, the presence of a massless Goldstone mode induces a logarithmic dependence on the vev.\\

As the loops evolve, they continuously lose energy through the emission of GWs or Goldstones, leading to a monotonic decrease in their initial length $l_i=\alpha t_i$ as,
\begin{align}
l(t) =  l_i - (\Gamma G \mu + \kappa) (t - t_i), \label{eq:length_shrink}
\end{align}
where $\Gamma\simeq 50$ \cite{Vilenkin:2000jqa, Vachaspati:1984gt}, $\alpha\simeq0.1$ \cite{Blanco-Pillado:2013qja,Blanco-Pillado:2017oxo}, $G$ is the Newton's constant, and $t_i$ is the time of loop formation. The shrinking rate is controlled by two contributions, $\Gamma G\mu$, associated with GW emission, and $\kappa$, associated with Goldstone production. For local strings, $\kappa = 0$, and loop decay is dominated by gravitational radiation. In contrast, global strings decay predominantly into Goldstone bosons, with an efficiency
$\kappa = \frac{\Gamma_{\rm Gold}}{2\pi} \log(v_{\rm B-L} t) \gg \Gamma G \mu$ ,
where  $\Gamma_{\rm Gold} \simeq 65$ \cite{Vilenkin:1986ku, Gorghetto:2021fsn}.

The total GW energy emitted by a loop can be decomposed into harmonics with instantaneous frequencies
\begin{equation}\label{eq:emitted_frequency}
    f_k = \frac{2k}{l(t)} = \frac{a(t_0)}{a(t)} f,
\end{equation}
where $k = 1, 2, 3, \dots, k_{\rm max}$, $f$ is the frequency observed today at $t_0$, and $a(t)$ is the scale factor. The total GW energy density is obtained by summing over all the $k$ modes leading to	
\begin{align}
\Omega_{\rm GW}(f) &=\sum_k\frac{1}{\rho_c}\cdot\frac{2k}{f}\cdot\frac{\mathcal{F}_\alpha \,\Gamma^{(k)}G\mu^2}{\alpha(\alpha+\Gamma G \mu + \kappa)} \times \nonumber\\
&\hspace{3em}  \int^{t_0}_{t_{\rm osc}}d\tilde{t} \, \frac{C_{\rm{eff}}(t_i)}{t_i^4}\left[\frac{a(\tilde{t})}{a(t_0)}\right]^5\left[\frac{a(t_i)}{a(\tilde{t})}\right]^3\Theta\left(t_i-\frac{l_*}{\alpha}\right)\Theta(t_i-t_{\rm osc}).
	\label{eq:master_eq_ready_to_use}
	\end{align}
Here, $\rho_c$ denotes the critical energy density of the universe, $\mathcal{F}_\alpha \simeq 0.1$ is an efficiency factor,  and $C_{\rm eff}(t_i)$  is the loop formation efficiency, computable
 from the velocity-dependent one-scale model \cite{Martins:1996jp, Martins:2000cs, Sousa:2013aaa, Auclair:2019wcv, Sousa:2020sxs}. The quantity $\Gamma_k=\Gamma k^{-\delta}/\zeta(\delta)$ parametrizes the fraction of GW power emitted into the $k$-th harmonic mode, where $\delta=4/3(5/3)$ for loops containing cusps (kinks)\cite{Damour:2001bk}. The integral in Eq. \eqref{eq:master_eq_ready_to_use} is regulated by two Heaviside functions, $\Theta\left(t_i-\frac{l_*}{\alpha}\right)\Theta(t_i-t_{\rm osc})$,
which imposes a high-frequency cut-off at $f_*$, beyond
which the GW spectrum exhibits a slope $f^{-1/3}$ when summed over a large number of modes. The quantity $t_{\rm osc}=\text{Max}\left[t_{\rm form},\,t_{\rm fric} \right]$ marks the epoch at which the motion of the string network ceases to be friction-dominated, or when loops that could have formed before the formation of the network are eliminated. The string network forms at the temperature of the $U(1)$ breaking phase transition,
\begin{equation}
    T_{\rm form}\simeq v_{\rm B-L},
\end{equation}
which corresponds, during radiation domination, to the formation time
\begin{equation} \label{tform}
    t_{\rm form} \simeq \left( \frac{90}{\pi^2 g_*(T_{\rm form})}\right)^{1/2} \frac{M_{\rm Pl}}{2 T_{\rm form}}. 
\end{equation}
At early times, the motion of strings is damped by interactions with the surrounding plasma. The temperature at which friction becomes inefficient can be estimated as \cite{Vilenkin:1991zk, Martins:1995tg, Martins:1996jp}
\begin{equation}\label{Tfric}
    T_{\rm fric} = \left( \frac{\pi^2 g_*(T_{\rm fric})}{90}\right)^{1/2}\frac{2 \mu}{\beta M_{\rm Pl} },
\end{equation}
where $\beta\sim \mathcal{O}(1)$ parametrizes the strength of particle–string interactions. The corresponding friction cutoff time follows from the standard time–temperature relation during radiation domination. Plugging Eq. \eqref{Tfric} into Eq. \eqref{tform}, we obtain the time of friction cut-off as
\begin{equation}
    t_{\rm fric}=\frac{2\pi v_{\rm B-L}}{\beta} \log (v_{\rm B-L} \tilde{t}_M) \left( \frac{g_*(T_{\rm fric})}{g_*(T_{\rm form})}\right)^{1/2}.
\end{equation}
The parameter $l_*$ represents a critical loop length above which GW emission dominates over particle production, as confirmed by high-resolution numerical simulations. For local strings, $l_\star$ corresponds to the onset of efficient massive particle emission from the string core and can be estimated as
\begin{equation}
    l_\star\sim \delta_w (\Gamma G\mu)^{-\gamma},
\end{equation}
where $\delta_w\sim (\sqrt{\lambda} v_{\rm B-L})^{-1}$ is the string width, with a scalar self coupling $\lambda$, and $\gamma=2(\gamma=1)$ for loops containing cusps (kinks) \cite{Matsunami:2019fss, Auclair:2019jip, Baeza-Ballesteros:2024sny}. For global strings, loop decay is instead dominated by Goldstone boson emission, and no sharp microscopic cutoff exists \cite{Baeza-Ballesteros:2023say}. In this case, $l_\star$ should be understood as an effective scale obtained by equating GW and Goldstone emission rates, giving 
\begin{equation}
    l_*\sim \kappa(\Gamma G \mu)^{-1},
\end{equation} 
which suppresses the GW contribution from small loops. Eq.~\eqref{eq:master_eq_ready_to_use} applies to both local and global strings, provided  Eq.~\eqref{eq:emitted_frequency} is used in conjunction with an appropriate choice of $\kappa$.

At high frequencies, under standard cosmological evolution, the GW spectrum from local strings is approximately flat \footnote{Ref.\cite{Schmitz:2024hxw} points out that cutoff or oscillatory features in the GW spectrum can arise in scenarios where GW emission from string loops becomes effective only after a delayed onset and is dominated by the fundamental oscillation mode. In the framework adopted here, GW emission is continuous once loops form and the spectrum is obtained by summing over harmonic modes, so such features are not generically expected.}, with an amplitude
\begin{equation}
    \Omega_{\rm std}^{\rm local} h^2 \simeq 15 \pi\Omega_r h^2 \, \Delta_T\, C_{\rm eff}^{\rm rad,l} \mathcal{F}_\alpha\left(\frac{\alpha G \mu}{\Gamma}\right)^{1/2},
\end{equation}
where $\Omega_{r}h^2 \simeq 4.2 \times 10^{-5}$\cite{ParticleDataGroup:2020ssz}. Small deviations from flatness may arise due to variations in the number of relativistic degrees of freedom, encapsulated in \cite{Gouttenoire:2019kij} by
\begin{align}
\label{eq:Delta_R}
\Delta_T \equiv \left( \frac{g_*(T)}{g_*(T_0)}\right)\left(\frac{g_{*s}(T_0)}{g_{*s}(T)} \right)^{4/3} .
\end{align}
In contrast, the high-frequency part of the GW spectrum from global strings is significantly suppressed and can be approximated as \cite{Gouttenoire:2019kij}
\begin{equation}
    \Omega_{\rm std}^{\rm global} h^2 \sim  90\Omega_r h^2 \, \Delta_T\,C_{\rm eff}^{\rm rad,g} \mathcal{F}_\alpha\left(\frac{\Gamma}{\Gamma_{\rm gold}}\right) \left(\frac{v_{\rm B-L}}{M_{\rm pl}}\right)^{4} \log^3\left( v_{\rm B-L} \tilde{t}_{\rm M}\right),
\end{equation}
where $\tilde{t}_M$ is the characteristic emission time that dominantly contributes to the present–day GW spectrum at a fixed observed frequency $f$. It arises from the interplay between loop shrinkage and redshifting, rather than from a maximum of the instantaneous GW power. For local strings, the GW power is constant in time, and this dominant time dependence cancels in high frequency regime. In contrast, for global strings GW emission is further suppressed by efficient Goldstone radiation, making early emission times particularly relevant. In practice, $\tilde{t}_M$ is obtained by maximizing the integrand in Eq. \eqref{eq:master_eq_ready_to_use} with respect to $\tilde{t}$, leading to 
\begin{equation}
    \tilde{t}_M =\frac{1}{t_0} \frac{4}{\alpha^2} \left( \frac{1}{f}\right)^2 \left( \frac{\alpha+\Gamma G \mu +\kappa}{\Gamma G \mu +\kappa}\right)^2.
\end{equation}
A detailed comparison between the GW signatures of local and global strings can be found in Refs.~\cite{Gouttenoire:2019kij,Ghoshal:2023sfa}.\\
\begin{figure}[H]
\centering
\includegraphics[width=0.49\linewidth]{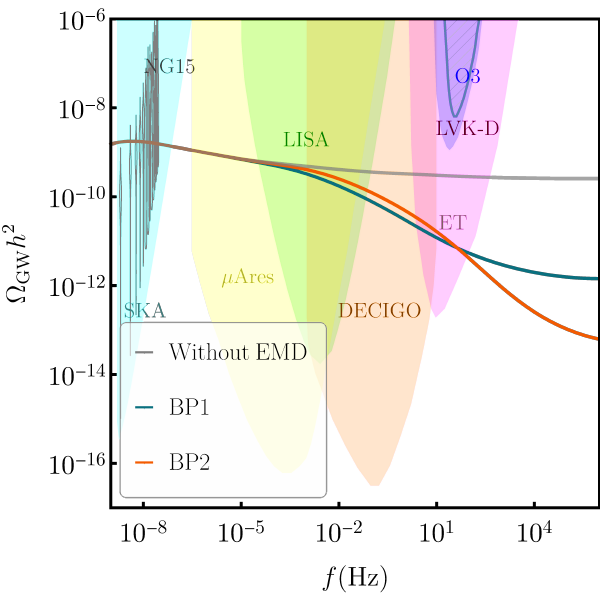}
\includegraphics[width=0.49\linewidth]{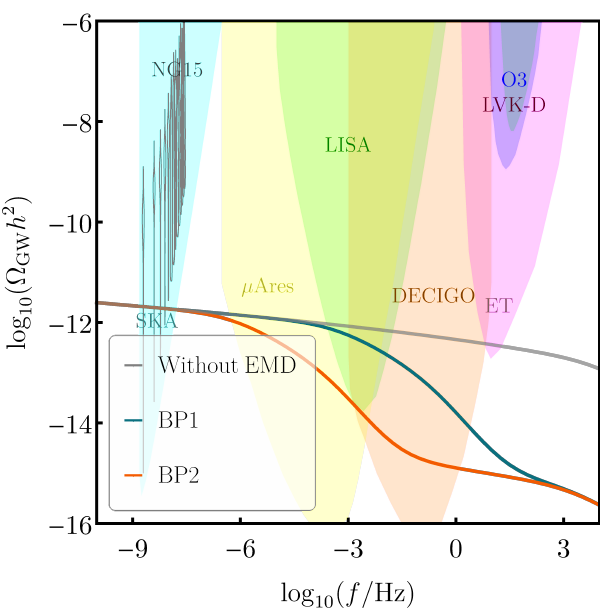} 
\caption{\it The GW spectral shapes for local strings (\textbf{left panel}) and global strings (\textbf{right panel}) are
shown with a brief period of early matter domination. The period of early matter domination is indicated by a kink and decline in the spectrum, whereas the spectrum is relatively flat for frequencies that correspond to periods of radiation
domination. Two features in the each benchmark point chosen, consisting of $f_{\rm dom}$ and $f_{\rm brk}$, determining the start and end of early matter dominations respectively (see Eq.\eqref{eq:fbreak},\eqref{turning_point_CS},\eqref{fdoml},\eqref{fdomg}), benchmarks are chosen in such a manner that in one benchmark (BP1) only one timescale that is the time of end of early matter domination, is observable, and in the other one (BP2), both the start and end of early matter domination timescales are observable (see Table \ref{tab:CSBP-table}). }
    \label{fig:CS benchmark}
\end{figure}
Following Refs. \cite{Ghoshal:2023sfa}, in scenarios with an early matter-dominated epoch, the otherwise flatter plateau of the GW spectrum takes a spectral turnover at a characteristic break frequency $f_{\rm brk}$. For cusp-dominated loop structures, and when summing over many harmonic modes, the spectrum above this break falls as $\Omega_{\rm GW}(f>f_{\rm brk})\propto f^{-1/3}$. The break frequency for local and global string networks can be estimated as,
\begin{equation}
    f_{\rm brk}^{\rm local}=6.32\times 10^{-3}\text{Hz} \left( \frac{T_{\rm brk}}{\rm GeV}\right) \left( \frac{0.1\times 50 \times 10^{-12}}{\alpha \Gamma G \mu}\right)^{1/2}  \left( \frac{g_{*}(T_{\rm brk})}{g_{*}(T_0)}\right)^{1/4} \label{eq:fbreak}
\end{equation}
and
\begin{equation} 
    f_{\rm brk}^{\rm global}= 8.9\times 10^{-7}\text{Hz} \left( \frac{T_{\rm brk}}{\rm GeV}\right) \left( \frac{0.1}{\alpha}\right)  \left( \frac{g_{*}(T_{\rm brk})}{g_{*}(T_0)}\right)^{1/4} 
    \label{turning_point_CS}
\end{equation}
where $T_{\rm brk}\equiv T_{\rm end}$ is the temperature at the end of early matter domination. Notably, $f_{\rm brk}^{\rm global}$ depends linearly on $T_{\rm brk}$ but is insensitive to the symmetry-breaking scale $v_{\rm B-L}$,  in contrast to the local string case. This distinction is illustrated in the GW spectra shown in Fig. \ref{fig:CS benchmark}.\\

However, in scenarios with a brief period of early matter domination, the GW spectrum exhibits a double-step feature that is more prominent for local strings. Interestingly, for local strings, prior to the flat plateau associated with early radiation domination, a characteristic knee feature arises due to loops that formed during the early radiation era but emit GWs during the early matter epoch \cite{Ghoshal:2023sfa}. The position of this knee can be well approximated as,
\begin{equation}
    f_{\rm knee}\simeq 6.07\times 10^3 {\:\rm Hz}\left(\frac{50\times 10^{-12}}{\Gamma G\mu}\right)\left( \frac{T_{\rm brk}}{\rm GeV}\right) \left( \frac{g_*(T_{\rm brk})}{g_*(T_0)}\right)^{1/4} ,
    \label{eq: fknee}
\end{equation}
with the GW amplitude given by
\begin{equation}
    \Omega_{\rm GW}^{\rm knee}\simeq \Omega_{\rm GW}^{\rm local}(T_{\rm brk}) \exp(-3N_e/4).
\end{equation}
Note that this feature is only visible if the loop lifetime is shorter than the early matter domination duration, which translates to
\begin{equation}
    N_e\lesssim 13.81+\frac{2}{3}\log\left[ \frac{50\times 10^{-12}}{\Gamma G\mu}\right].
\end{equation}
The characteristic high-frequency turning point for local strings occurs at
\begin{equation}
    f_{\rm dom}^{\rm local}\simeq f_{\rm brk}^{\rm local} \exp(3N_e),
    \label{fdoml}
\end{equation}
where the GW amplitude can be well approximated as
\begin{equation}
    \Omega_{\rm GW}^{\rm local}(f_{\rm dom}^{\rm local})\simeq \Omega_{\rm GW,std}^{\rm local}(f_{\rm dom}^{\rm local})\exp(-3N_e).
\end{equation}
On the other hand, for global strings, the characteristic high-frequency turning point occurs at frequency
\begin{equation}
    f_{\rm dom}^{\rm global}\simeq f_{\rm brk}^{\rm global} \exp(3N_e)\left(\frac{\log\left[ (5.6\times 10^{30}) \left( \frac{v_{\rm B-L}}{10^{15} \rm GeV}\right)\left( \frac{10^{-3}\rm Hz}{f_{\rm brk}^{\rm global}}\right)^2\right]}{\log\left[ (5.6\times 10^{30}) \left( \frac{v_{\rm B-L}}{10^{15} \rm GeV}\right)\left( \frac{10^{-3}\rm Hz}{f_{\rm dom}^{\rm global}}\right)^2\right]}\right)^9,
    \label{fdomg}
\end{equation}
where the GW amplitude can be well approximated as
\begin{equation}
    \Omega_{\rm GW}^{\rm global}(f_{\rm dom}^{\rm global})\simeq \Omega_{\rm GW,std}^{\rm global}(f_{\rm dom}^{\rm global})\exp(-3N_e). \label{eq:fdom}
\end{equation}
\begin{table}[]
\begin{tabular}{|l||ll|ll|}
\hline
\textbf{BP} & \multicolumn{2}{l|}{\textbf{local CS ($10^{14}$ GeV)}}         & \multicolumn{2}{l|}{\textbf{global CS ($10^{15}$ GeV)}}        \\ \hline
            & \multicolumn{1}{l|}{$T_{\rm dom}$ (GeV)} & $T_{\rm dec}$ (GeV) & \multicolumn{1}{l|}{$T_{\rm dom}$ (GeV)} & $T_{\rm dec}$ (GeV) \\ \hline
1           & \multicolumn{1}{l|}{$10^1$}              & $10^{-1}$           & \multicolumn{1}{l|}{$10^1$}              & $10^{0.2}$          \\ \hline
2           & \multicolumn{1}{l|}{$10^3$}              & $10^0$              & \multicolumn{1}{l|}{$10^3$}              & $10^2$              \\ \hline
\end{tabular}
\caption{\it The benchmark cases presented in Fig.\ref{fig:CS benchmark} for the RHN sourced early matter domination.}
\label{tab:CSBP-table}
\end{table}

\subsubsection{Gravitational Wave Detectors} 
In the GW spectrum plots, in Fig. \ref{fig:CS benchmark}, we display the power-law integrated sensitivity curves for a myriad of ongoing and future GW experiments. They can be grouped as: 
\begin{itemize}
    \item \textbf{Ground-based interferometers:} These detectors, such as \textsc{LIGO}/\textsc{VIRGO} \cite{LIGOScientific:2016aoc,LIGOScientific:2016sjg,LIGOScientific:2017bnn,LIGOScientific:2017vox,LIGOScientific:2017ycc,LIGOScientific:2017vwq}, a\textsc{LIGO}/a\textsc{VIRGO} \cite{LIGOScientific:2014pky,VIRGO:2014yos,LIGOScientific:2019lzm}, \textsc{AION} \cite{Badurina:2021rgt,Graham:2016plp,Graham:2017pmn,Badurina:2019hst}, \textsc{Einstein Telescope (ET)} \cite{Punturo:2010zz,Hild:2010id}, and \textsc{Cosmic Explorer (CE)} \cite{LIGOScientific:2016wof,Reitze:2019iox}, use interferometric techniques on the Earth's surface to detect gravitational waves.
    
    \item \textbf{Space-based interferometers:} Space-based detectors like \textsc{LISA} \cite{Baker:2019nia}, \textsc{BBO} \cite{Crowder:2005nr,Corbin:2005ny,Cutler:2009qv}, \textsc{DECIGO}, \textsc{U-DECIGO} \cite{Seto:2001qf,Yagi:2011wg}, \textsc{AEDGE} \cite{AEDGE:2019nxb,Badurina:2021rgt}, and \textsc{$\mu$-ARES} \cite{Sesana:2019vho} are designed to detect gravitational waves from space, offering different advantages over ground-based counterparts.
    
    \item \textbf{Recasts of star surveys:} Monitoring of star surveys like \textsc{GAIA}/\textsc{THEIA} \cite{Garcia-Bellido:2021zgu} utilize astrometric data from stars can indirectly infer the presence of gravitational wave signals.
    
    \item \textbf{Pulsar timing arrays (PTA):} PTA experiments like \textsc{SKA} \cite{Carilli:2004nx,Janssen:2014dka,Weltman:2018zrl}, \textsc{EPTA} \cite{EPTA:2015qep,EPTA:2015gke}, and \textsc{NANOGRAV} \cite{NANOGRAV:2018hou,Aggarwal:2018mgp,NANOGrav:2020bcs} use precise timing periodicity measurements of pulsars to detect gravitational wave signatures.
    
    
\end{itemize}

\medskip

\medskip
\subsection{Gravitational Wave Tests of Long-Lived right-handed Neutrinos}
\begin{figure}[H]
\centering
\includegraphics[width=0.49\linewidth]{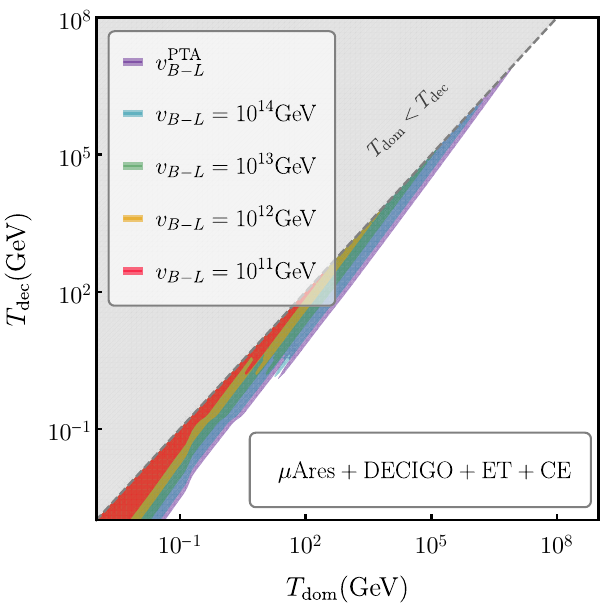} 
\includegraphics[width=0.49\linewidth]{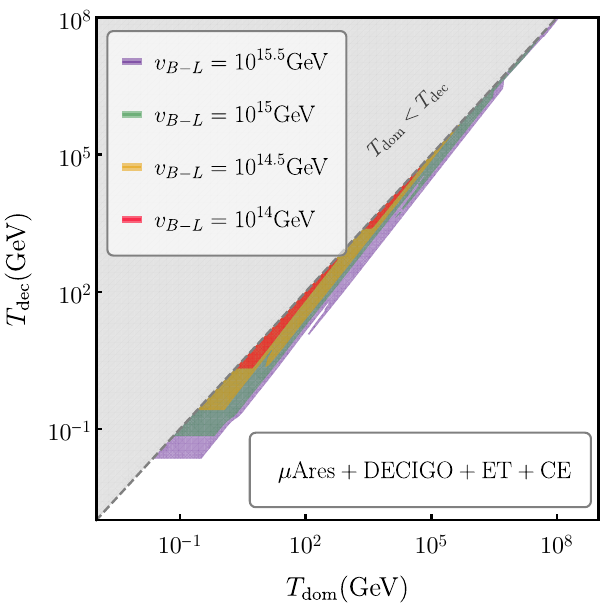} 
\caption{\it The phenomenologically interesting parameter space on $T_{\rm dom}$ vs $T_{\rm dec}$ plane is essentially determined by whether the characteristic frequencies $f_{\rm brk}$, and $f_{\rm dom}$
, associated with the end and onset of an early matter-dominated era, can be simultaneously probed. Detectability is assessed using the combined sensitivity envelope of future GW detectors ($\mu$Ares, DECIGO, ET, CE, etc.), requiring that both frequencies lie within the sensitivity reach across the corresponding frequency bands (see text for details). For local (\textbf{left panel}) and global (\textbf{right panel}) $B-L$ strings.}
    \label{fig:kappaThermal1}
\end{figure}

\begin{figure}[H]
\centering
\includegraphics[width=0.49\linewidth]{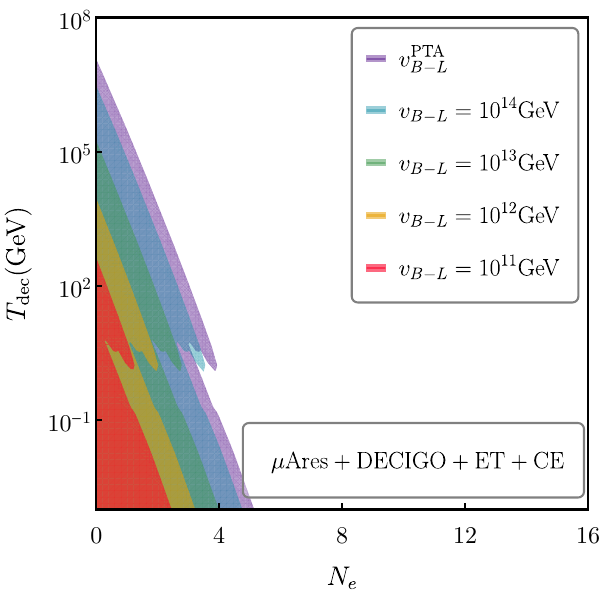} 
\includegraphics[width=0.49\linewidth]{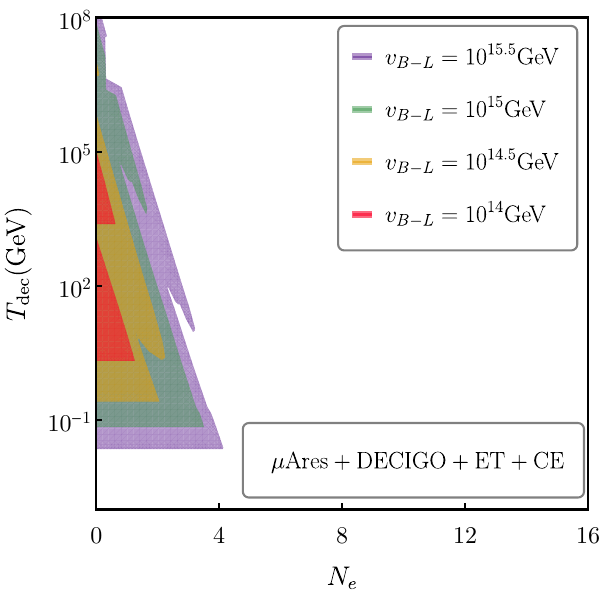} 
\caption{\it Plot showing the impact of early matter domination duration on GW detectability. For local (\textbf{left panel}) and global (\textbf{right panel}) $B-L$ strings.}
    \label{fig:kappaThermal2}
\end{figure}

By correlating the results for the detectability of the modified gravitational wave spectrum from cosmic strings, we identify the experimentally testable regions of the temperature parameter space. The testable regions by future experiments in the temperature plane are shown in Figure \ref{fig:kappaThermal1}. Figure \ref{fig:kappaThermal2} shows the testable regions characterised by the end of early matter domination and by its duration, parametrised by $N_e$.

Although the detection of a break or turning point in the SGWB (see Fig. \ref{fig:CS benchmark} for understanding the deviation from the standard scale-invariant GW spectrum) from cosmic strings would provide a powerful diagnostic of the Universe's evolutionary history, such a break is degenerate in its origin, potentially indicating an extended era of early matter domination sourced by metastable species or exotic states \cite{Gouttenoire:2019kij,Ghoshal:2023sfa}, the imprint of a supercooled phase transition \cite{Gouttenoire:2019kij,Ferrer:2023uwz,Datta:2025owx}, or a high-frequency cut-off from cusp-collision dynamics \cite{Damour:2001bk,Auclair:2019jip}. A particularly compelling and unique signature, however, is predicted for a transient, brief matter-dominated era. This scenario generates a sharp, step-like feature demarcated by two observable kinks in the GW spectrum, one corresponding to the onset of early matter domination and the other to its end (see Fig. \ref{fig:CS benchmark} and Sec.\ref{subCS} for details). To assess the full phenomenological potential of this scenario, 
our analysis adopts an optimistic detection framework. Our primary goal is to identify both turning points in the GW spectrum associated with the onset and end of an early matter-dominated era, which typically span several decades in frequency and cannot be simultaneously probed by a single detector. We therefore define a ‘combined sensitivity’ as the envelope of the projected sensitivities of future GW observatories such as $\mu$Ares, LISA, DECIGO, CE and ET, effectively treating them as a single broadband probe across a wide frequency range with a signal-to-noise ratio (SNR) defined as \cite{Maggiore:1999vm,Schmitz:2020syl},
\begin{equation}
    {\rm SNR} =\sqrt{t_{\rm obs}\int_{f_{\rm min}}^{f_{\rm max}}df\left( \frac{\Omega_{\rm GW}(f)}{\Omega_{\rm noise}(f)}\right)^2},
\end{equation}
with $t_{\rm obs}=10$ years, where $\Omega_{\rm noise}$ represents the noise curve of a given experiment, and $f_{\rm max}(f_{\rm min})$ are the maximum (minimum) accessible frequency. This approach allows us to define the parameter space in which both critical kinks are simultaneously resolvable, providing a clear target for future missions. We chose SNR $ \geq 1$ as our detection threshold for the detection of the characteristic features in the GW spectrum, which we describe in detail below.

\begin{figure}[H]
\centering
\includegraphics[width=0.9\linewidth]{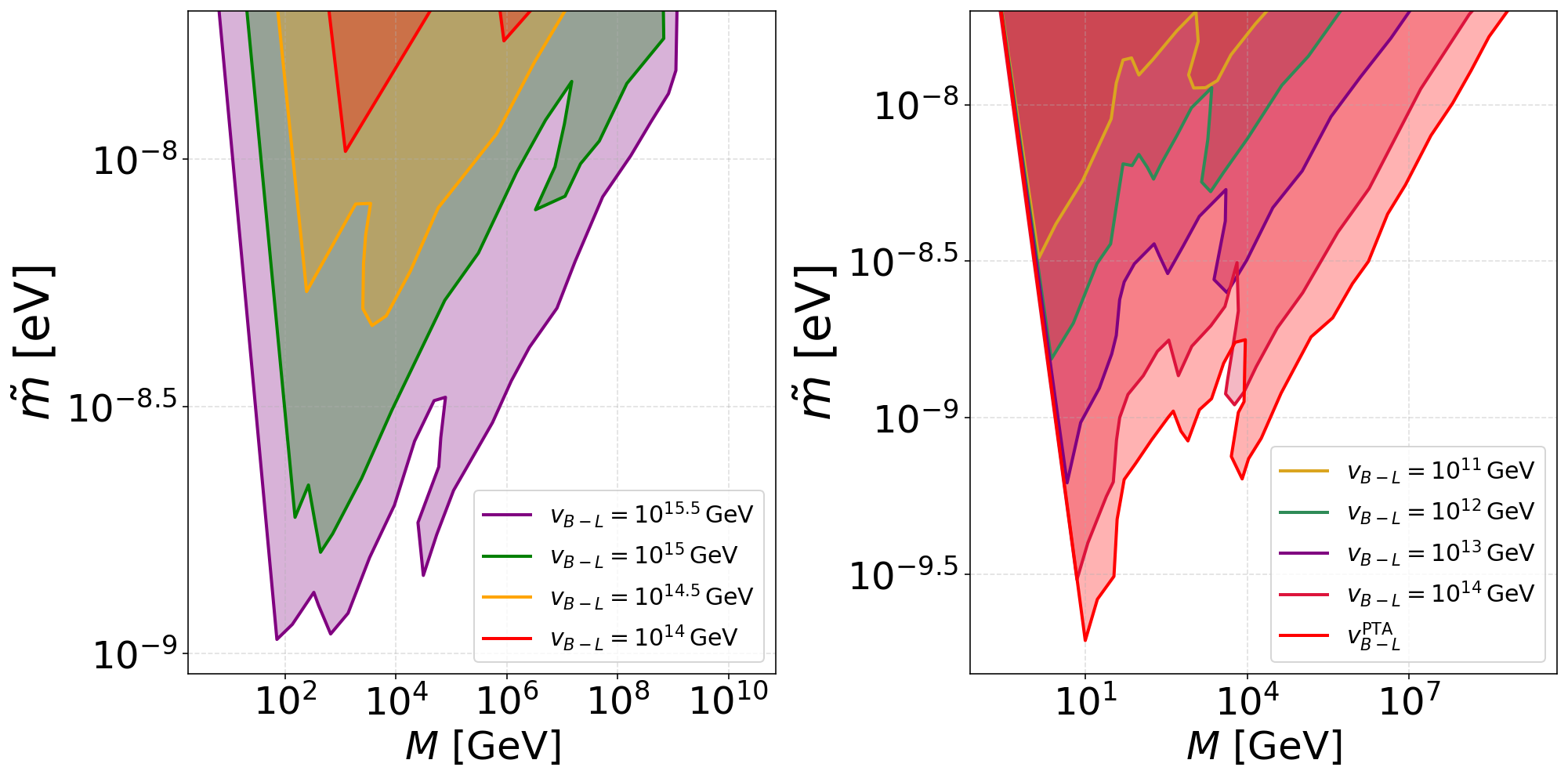} 
\caption{\it Detectable regions of the seesaw parameter space inferred from the gravitational-wave background, where both the breaking and early-matter domination frequencies are determined using the best-fit relations (Eqs.~\ref{eq: Tdom}, \ref{eq: Tend}). Detectability is defined by requiring that both characteristic frequencies are accessible within the combined sensitivity envelope of future GW detectors}. The panels show global (left) and local (right) symmetry cases shown in the mass plane. These regions allow us to determine the mass of the right-handed neutrino and the effective neutrino mass related to the EMD caused by the right-handed neutrino. 
    \label{fig:Mass experiment bound}
\end{figure}

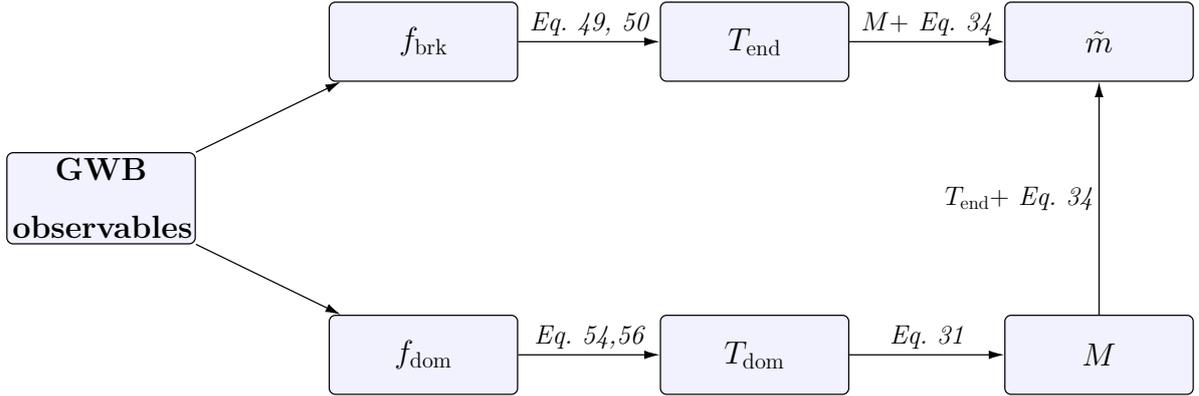
\begin{figure}[t]
    \centering
    \resizebox{0.95\linewidth}{!}{%
    \begin{tikzpicture}[
        node distance=3.0cm and 2.6cm,
        box/.style={
            rectangle, rounded corners, draw=black, fill=blue!5, thick,
            text centered, text width=4.0cm, minimum height=1.8cm,
            font=\bfseries\LARGE    
        },
        arrow/.style={thick, -{Latex[length=3.5mm,width=2mm]}},
        eqnlabel/.style={midway, above, sloped, font=\Large\itshape},
        eqnlabelV/.style={pos=0.5, left, font=\Large\itshape}
    ]

    \node[box] (gwb) {GWB\\ observables};

    \node[box, above right=1.6cm and 3.cm of gwb] (fbrk) {$f_{\mathrm{brk}}$};
    \node[box, below right=1.6cm and 3.cm of gwb] (fdom) {$f_{\mathrm{dom}}$};

    \node[box, right=3.2cm of fdom] (tdom) {$T_{\mathrm{dom}}$};
    \node[box, right=3.2cm of fbrk] (tend) {$T_{\mathrm{end}}$};

    \node[box, right=3.5cm of tdom] (mass) {$M$};
    \node[box, right=3.5cm of tend] (mtilde) {$\tilde{m}$};

    \draw[arrow] (gwb) -- (fbrk);
    \draw[arrow] (gwb) -- (fdom);

    \draw[arrow] (fdom) -- node[eqnlabel] {Eq.~\ref{fdoml},\ref{fdomg}} (tdom);
    \draw[arrow] (tdom) -- node[eqnlabel] {Eq.~\ref{eq: Tdom}} (mass);

    \draw[arrow] (fbrk) -- node[eqnlabel] {Eq.~\ref{eq:fbreak},~\ref{turning_point_CS}} (tend);

    \draw[arrow] (tend.east) -- node[eqnlabel] {$M +$ Eq.~\ref{eq: Tend}} (mtilde.west);
    \draw[arrow] (mass.north) -- node[eqnlabelV] {$T_{\mathrm{end}} +$~Eq.~\ref{eq: Tend}} (mtilde.south);

    \end{tikzpicture}%
    }

    \caption{
        \it Flow chart from a gravitational-wave background (GWB) to seesaw parameters.
        A stochastic GWB can carry two key frequencies: $f_{\mathrm{dom}}$, marking the onset of early matter domination, and $f_{\mathrm{brk}}$, corresponding to the return to radiation domination.
        From $f_{\mathrm{dom}}$, one infers the temperature of domination $T_{\mathrm{dom}}$ via Eqns.~(\eqref{fdoml},\eqref{fdomg}), which determines the right-handed neutrino mass $M$ through Eq.~\ref{eq: Tdom}.
        The break frequency $f_{\mathrm{brk}}$ gives the decay temperature $T_{\mathrm{end}}$ via Eqs.~(\eqref{fdoml},\eqref{fdomg}).
        Finally, combining $M$ and $T_{\mathrm{end}}$ using Eq.~\ref{eq: Tend} determines the effective neutrino mass $\tilde{m}$, providing a direct link between gravitational-wave observables and the seesaw parameters.
  }
    \label{fig:GWB_inference_chain}
\end{figure}

Using our best fit formulas for the onset and termination of early matter domination in terms of the seesaw physics parameters $M$ and $\tilde m$, these detectable ranges can be trivially converted to detectable regions of the mass parameter spaces instead of temperature. This detectable parameter space is shown in Figure \ref{fig:Mass experiment bound}, while the overall procedure correlating gravitational-wave observables to the underlying seesaw parameters is illustrated in Fig.~\ref{fig:GWB_inference_chain}.
In Fig. \ref{fig:kappaThermal1}, a monotonic suppression of the SGWB amplitude is observed with decreasing vev. This suppression progressively erodes the sensitivity of GW detectors across their maximum operational frequency bands. Beyond a critical threshold in amplitude, the signal effectively becomes observationally inaccessible. Consequently, the combined reach of the detector network fails to capture the complete spectral feature across a significant portion of the parameter space. This dynamic results in the emergence of isolated, island-like regions of detectability, which are confined to progressively lower vevs.
The Fig. \ref{fig:Mass experiment bound} clearly shows that the gravitational waves from global cosmic strings can probe right-handed neutrino masses across a range of nine orders of magnitude, spanning from approximately $M \sim 10\,\mathrm{GeV}$ up to $M \sim 10^{10}\,\mathrm{GeV}$, with sensitivity down to effective masses of order $\tilde m \sim 10^{-9}\,\mathrm{eV}$. This broad reach in $M$ arises from varying the Yukawa couplings that relate the right-handed neutrino masses to the underlying symmetry-breaking scale, $M=y_N v_{\rm B-L}$, while the gravitational-wave signal itself is primarily controlled by $v_{\rm B-L}$. Consequently, sizeable hierarchies between $v_{\rm B-L}$ and $M$ are both technically natural and phenomenologically relevant. Nonetheless, local strings extend this reach considerably, covering right-handed neutrino masses from $M \sim 0.1\,\mathrm{GeV}$ scale to $10^{9}\,\mathrm{GeV}$, while probing effective mass as small as $\tilde m \sim 10^{-10}\,\mathrm{eV}$. Symmetry breaking with larger vevs leads to a higher amplitude of the gravitational-wave background, while shifting the characteristic frequencies to much lower values, making them easier to detect. These results indicate that gravitational wave backgrounds from cosmic strings can explore a vast range of parameter space corresponding to such exotic early matter domination by right-handed neutrinos\footnote{If these strings are meta-stable \cite{meta1, meta2, meta5, meta7}, the PTA bound can be relaxed, allowing us to explore nearly the entire viable leptogenesis parameter space. Interestingly, in this case, an EMD is naturally motivated \cite{Datta:2024bqp} to avoid the LIGO O3's null result at higher frequencies.}. In particular, the sensitivity to such small values of $\tilde m$ demonstrates that even extremely weakly coupled right-handed neutrinos, corresponding to lifetimes far beyond laboratory reach, leave an imprint that could be accessible to future GW detectors. Since the seesaw requires heavy Majorana neutrinos, mapping out this space is directly tied to testing the Majorana nature of neutrinos. Even laboratory searches, such as neutrino-less double beta decay\footnote{From the experimental constraints of neutrinoless double $\beta$-decay from KATRIN, the direct neutrino mass measurement gives \cite{KATRIN:2021uub} (see also Refs. \cite{Dolinski:2019nrj,Gomez-Cadenas:2011oep}), \mbox{$m_{\nu} \leq 0.8$ eV} and future sensitivity can reach up to \mbox{$m_{\nu} \leq 0.2$ eV}.}, often regarded as hallmark laboratory tests of the seesaw mechanism, only constrain low‑energy effective parameters of the light neutrinos and cannot access the effective mass $\tilde{m}$ associated with heavy right‑handed states. When such states lie at scales beyond the reach of direct collider searches, gravitational waves may provide the only indirect and complementary probe of this crucial parameter of the type‑I seesaw. The impact of the presence of heavy RHN on the SM Higgs vacuum stability \cite{Ipek:2018sai} has also been examined, but the associated bound is significantly higher than any sterile neutrino masses considered here and is therefore not constraining for our analysis. Another commonly discussed concern is Yukawa perturbativity. In our case, however, the effective neutrino masses are extremely small, ensuring that the Yukawa couplings remain well within the perturbative regime. For instance, even for $M=10^{12}$ GeV, the perturbativity bound corresponds to $\tilde m_{\rm max}\lesssim 400$ eV, far above the values considered here.  

\medskip

\section{Primordial Gravitational Wave Tests of Leptogenesis}
\label{sec:leptogenesis}
The origin of the matter–antimatter asymmetry remains one of the central unresolved problems in cosmology and particle physics. The baryon asymmetry is measured independently from the cosmic microwave background \cite{Planck2018} and Big Bang Nucleosynthesis \cite{ParticleDataGroup:2020ssz} 
\begin{equation}
    Y_B=\frac{n_B}{s}= 8.87\times 10^{-11}\ .
\end{equation}
Leptogenesis \cite{Fukugita:1986hr, Luty:1992un} offers a compelling and theoretically consistent framework that reproduces the observed baryon asymmetry: CP-violating decays of heavy right-handed neutrinos consistent with the Sakharov conditions \cite{Sakharov:1967dj} generate a lepton asymmetry that electroweak sphalerons subsequently convert into a baryon asymmetry \cite{Luty:1992un, Giudice_2004, Covi_1996, Buchm_ller_2005}. In the parameter regime considered here, where the Universe experiences a transient matter-dominated phase due to long-lived sterile neutrinos, the dynamics of leptogenesis are modified by both the altered expansion rate and subsequent entropy dilution. The goal of this section is to identify the regions in the $(M_1, \tilde m_1)$ parameter space that yield the observed baryon asymmetry and are potentially testable through cosmological signatures. The key parameter governing successful leptogenesis in both the hierarchical and near-resonant regimes is the right-handed neutrino mass \(M_1\). As \(M_1\propto v_{\rm B-L}\) originates from the spontaneous breaking of the \(U(1)_{B-L}\) symmetry, the generation of the baryon asymmetry can ultimately be traced back to the dynamics of \(U(1)_{B-L}\) breaking. The scale of this breaking also determines whether the gravitational-wave background can probe the period of early matter domination through the detection of the characteristic frequencies \(f_{\rm brk}\) and \(f_{\rm dom}\).

\subsection{The CP asymmetry}
\label{sec: CP asymmetry}
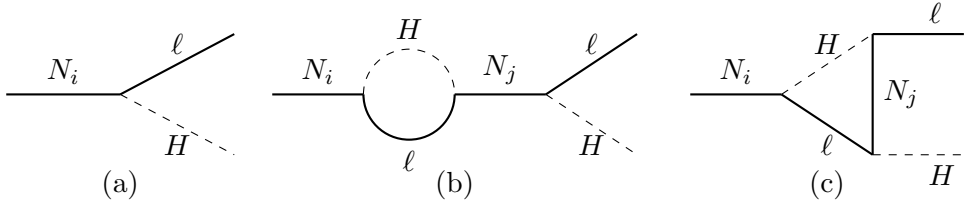
\begin{figure}[h!]
\centering
\begin{tikzpicture}[scale=1, baseline=(current bounding box.center)]

  \coordinate (Ni1) at (0,0);
  \coordinate (v1) at (1.5,0);
  \coordinate (ell1) at (3,0.8);
  \coordinate (phi1) at (3,-0.8);
  \draw[thick] (Ni1) -- (v1) node[midway, above] {\( N_i \)};
  \draw[thick] (v1) -- (ell1) node[midway, above ] {\( \ell \)};
  \draw[dashed] (v1) -- (phi1) node[midway, below ] {\( H \)};
  \node at (1.5, -1.2) {(a)};

  \begin{scope}[xshift=3.5cm]
    \coordinate (Ni2) at (0,0);
    \coordinate (loopL) at (1.2,0);
    \coordinate (loopR) at (2.4,0);
    \coordinate (v2) at (3.6,0);
    \coordinate (ell2) at (4.8,0.8);
    \coordinate (phi2) at (4.8,-0.8);
    \draw[thick] (Ni2) -- (loopL) node[midway, above] {\( N_i \)};
    \draw[thick] (loopR) -- (v2) node[midway, above] {\( N_j \)};
    \draw[thick] (v2) -- (ell2) node[midway, above ] {\( \ell \)};
    \draw[dashed] (v2) -- (phi2) node[midway, below ] {\( H \)};
    \draw[dashed] (loopL) arc[start angle=180, end angle=0, x radius=0.6cm, y radius=0.6cm] node[midway, above] {\( H \)};
    \draw[thick] (loopL) arc[start angle=-180, end angle=0, x radius=0.6cm, y radius=0.6cm] node[midway, below] {\( \ell \)};
    \node at (2.4, -1.2) {(b)};
  \end{scope}

  \begin{scope}[xshift=9cm]
    \coordinate (Ni3) at (0,0);
    \coordinate (vtx) at (1.2,0);
    \coordinate (Nj) at (2.4,0);
    \coordinate (ellOut) at (3.6,0.8);
    \coordinate (phiOut) at (3.6,-0.8);
    \coordinate (ellLoop) at (2.4,-0.8);
    \coordinate (phiLoop) at (2.4,0.8);
    \draw[thick] (Ni3) -- (vtx) node[midway, above] {\( N_i \)};
    \draw[thick] (vtx) -- (ellLoop) node[midway, below] {\( \ell \)};
    \draw[dashed] (vtx) -- (phiLoop) node[midway, above] {\( H \)};
    \draw[thick] (ellLoop) -- (phiLoop) node[midway, right] {\( N_j \)};
    \draw[thick] (phiLoop) -- (ellOut) node[midway, above right] {\( \ell \)};
    \draw[dashed] (ellLoop) -- (phiOut) node[midway, below right] {\( H \)};
    \node at (1.8, -1.2) {(c)};
  \end{scope}

\end{tikzpicture}
\caption{\it Feynman diagrams contributing to the CP asymmetry: (a) tree-level, (b) self-energy, and (c) vertex diagrams.}
\label{fig:cpasymmetrydiagrams}
\end{figure}

In leptogenesis, a lepton asymmetry is generated through the $CP$-violating decays of heavy right-handed neutrinos into a Higgs boson and a lepton at one loop order. The relevant decay processes are depicted in Figure \ref{fig:cpasymmetrydiagrams}.
\noindent The $CP$ violation arises from interference between tree-level and one-loop decay diagrams, producing different decay rates for right-handed neutrino decays into Higgs and leptons, $N \rightarrow H L$ and $N \rightarrow H^\dag \bar{L}$. This CP violation is quantified by the parameter $\epsilon$, defined as:
\begin{equation}
    \epsilon_i = \frac{\Gamma(N_i \rightarrow L H) - \Gamma(N_i \rightarrow \bar{L} H^\dagger)}{\Gamma(N_i \rightarrow L H) + \Gamma(N_i \rightarrow \bar{L} H^\dagger)},
\end{equation}
where $\Gamma$ represents the decay rate of the processes. The precise value of $\epsilon$ depends on the masses of the right-handed neutrinos and their Yukawa couplings \cite{Covi_1996, Buchm_ller_2005, Giudice_2004, Di_Bari_2012}
\begin{equation}
    \epsilon_i = \frac{1}{8 \pi}\sum_{j\neq i}\frac{{\rm Im}[(y^\dagger y)_{ij}^2]}{(y^\dagger y)_{ii}}  f\left(\frac{M_j^2}{M_i^2}\right)\ .
    \label{eq: epsilon}
\end{equation}
where the function $f(x)$ captures the dependence on the mass ratio of the right-handed neutrinos
\begin{equation}
    f(x) = \sqrt{x} \left[(1 + x) \log\left(\frac{1 + x}{x}\right) - \frac{2 - x}{1 - x}\right].
\end{equation}
From this expression, we observe that $\epsilon$ becomes large when the masses of two right-handed neutrinos are nearly degenerate, $M_i \approx M_j$, leading to the phenomenon called resonant leptogenesis \cite{Pilaftsis_2004}. This scenario requires a refined treatment of the self-energy contribution, which dominates in this limit and develops a regulated enhancement. Various prescriptions for regulating the divergent behaviour have been proposed in the literature \cite{Klaric_2021, Pilaftsis_2004, Riotto_2007, Garbrecht_2014, Garny_2013, Anisimov_2006}. Owing to this ambiguity, in the present study, we refrain from analysing this regime. To this end, we shall impose the condition that the mass splitting is much greater than the decay rates of the right-handed neutrinos \cite{Moffat_2018}. Enhancements can occur through other mechanisms such as soft leptogenesis \cite{Boubekeur:2002jn, D_Ambrosio_2003}; however, they require extending the model with additional supersymmetric particles and soft-breaking terms, which goes beyond the minimal framework considered here.
\subsection{Entropy Dilution and the Boltzmann Equations}
\begin{figure}[H]
    \centering
    \includegraphics[width=0.49\linewidth]{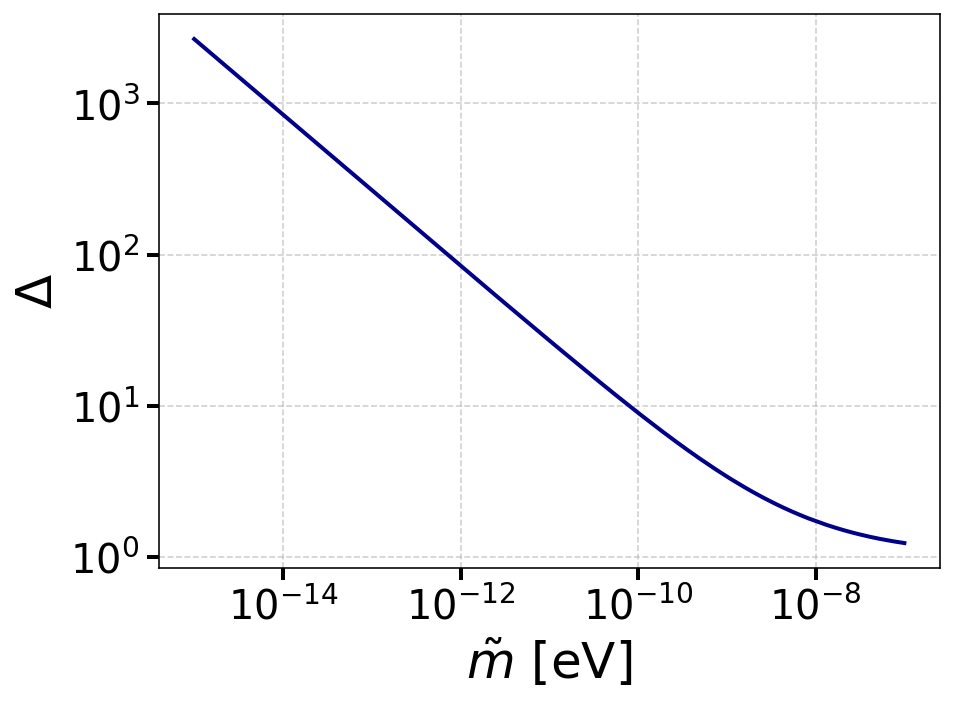}
    \includegraphics[width=0.49\linewidth]{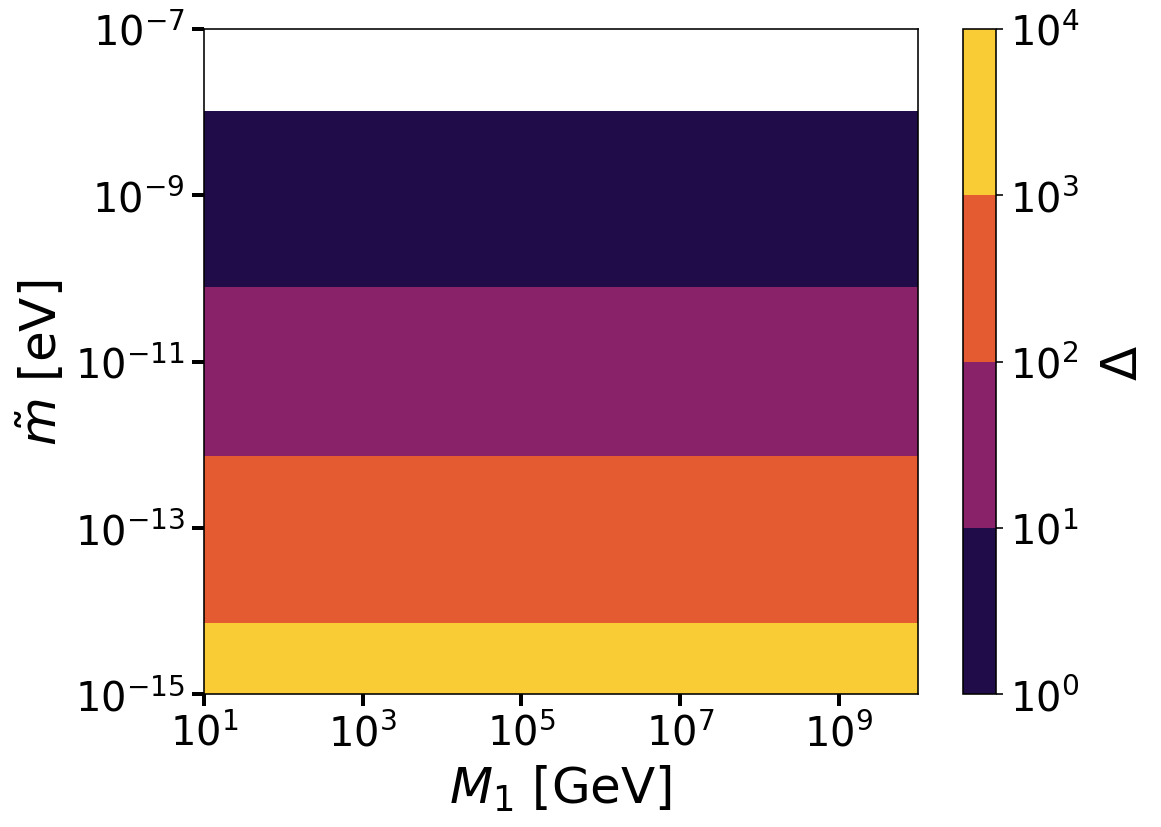}
    \caption{\it Entropy dilution as a function of seesaw parameters. 
    \textbf{Left:} Dependence on the effective neutrino mass. A smaller effective mass extends the period of early matter domination, leading to the right-handed neutrino energy density dominating the energy budget to a larger extent. This gives a larger entropy dump when the right-handed neutrinos eventually decay.
    \textbf{Right:} dependence on both $M_1$ and $\tilde m$ showing entropy dilution is independent of right-handed neutrino mass.}
    \label{fig:entropy scan}
\end{figure}

The dynamics of leptogenesis in a matter-dominated background differ significantly from the standard radiation-dominated scenario. In particular, the expansion rate is modified, and the entropy injection from right-handed neutrino decays dilutes any generated asymmetry. The sudden transfer of their non-relativistic energy density $\rho_N$ into radiation $\rho_R$ produces a significant increase in the comoving entropy. This entropy injection dilutes any pre-existing comoving number densities, including the baryon or lepton asymmetry generated during or before the decay. For instance, if an asymmetry $Y_B$ is produced while the Universe is dominated by $\rho_N$, the final observed baryon asymmetry after decay is suppressed by a factor $\Delta^{-1}$, where $\Delta$ is the entropy injection factor. For relativistic radiation, the entropy density scales as $s \propto \rho_R^{3/4}$, so taking the assumption of instantaneous decays, the entropy injection is parametrised by,
\begin{equation}
\Delta \equiv \frac{s_{\text{after}}}{s_{\text{before}}} 
= \left( \frac{\rho_R + \rho_N}{\rho_R} \right)^{3/4} 
= \left( 1 + \frac{\rho_N}{\rho_R} \right)^{3/4}= \left( 1 + \frac{T_{\rm dom}}{T_{\rm dec}} \right)^{3/4},
\label{eq:Delta}
\end{equation}
where $\rho_R$ and $\rho_N$ are evaluated just before decays begin, and in the last step we have used Eq. \ref{eq: rho to T}. A large $\rho_N / \rho_R$ ratio corresponds to a strong matter-dominated era and hence a large $\Delta$, which can drastically suppress the final baryon asymmetry. We performed a scan, for each point $(M_1,\tilde m_1)$, we took the maximum right-handed neutrino to radiation energy ratio as the input to the entropy dilution. We found that the entropy dilution is independent of right-handed neutrino mass and is monotonically decreasing with effective neutrino mass. This is shown in figure \ref{fig:entropy scan}.
This has the best-fit formula,
\begin{equation}
    \Delta = \left( 1 + 3.726 \times 10^{-6} \,\left(\frac{\tilde m}{\rm eV}\right)^{-0.67} \right)^{3/4}.
\end{equation}
The entropy dilution is heavily related to the duration of early matter domination, since a prolonged phase enables right-handed neutrinos to overtake the energy content more substantially, which amplifies the entropy injection when they eventually decay.\\
To calculate the baryon asymmetry, we solve the Boltzmann equations for the lepton asymmetry as well as for the other variables. The asymmetry equation has two contributions: a source term, which generates the asymmetry, and a washout term, which accounts for inverse decays and $\Delta L=2$ scatterings \cite{Fong_2012, Ulysses, Ulysses2}. 
\begin{equation}
\begin{aligned}
    \frac{dY_N}{dz}&=-D(z)\Big(Y_N-Y_N^{\rm eq}(z)\Big) \\
    \frac{dY_R}{dz}&=2 D(z)\Big(Y_N-Y_N^{\rm eq}(z)\Big) \\
    \frac{dY_{B-L}}{dz}& = \epsilon D(z) \Big(Y_N - Y_N^\text{eq}(z)\Big) - W(z) Y_{B-L}\ .
\end{aligned}
\end{equation}
$D(z)$ is the decay term in Eq. \ref{eq: D}, the CP asymmetry parameter, $\epsilon$, was given in Eq. \ref{eq: epsilon}. $W(z)$ describes the washout. In the large $z$ limit, the equilibrium number density is exponentially suppressed \cite{Moffat_2018, Ulysses, Ulysses2},
\begin{equation}
    Y_N^{\rm eq}(z)=\frac{45\,g_N}{4\pi^4 g_*}\,z^2 K_2(z)\xrightarrow[z\gg 1]\ 0
\end{equation}
and therefore, we have a vanishing washout term \cite{Moffat_2018, Ulysses, Ulysses2}
\begin{equation}
    W(z)=\frac{1}{2}D(z)\frac{Y_N^{\rm eq}}{Y_L^{\rm eq}}\rightarrow 0\ .
\end{equation}
In this regime, there is negligible erasure of the produced asymmetry. Moreover, with no efficient washout interactions, flavour effects \cite{Nardi:2006fx, Abada_2006, Antusch_2006, Blanchet_2007, DeSimone:2006nrs, Cirigliano_2010, Simone_2007, Racker_2012, Roshan:2025mwl} are absent: the asymmetry generated in each lepton flavour evolves identically, so the single unflavoured Boltzmann equation is the appropriate equation to evolve the asymmetry. Temperature effects \cite{Giudice_2004} can also be omitted as leptogenesis in the regime $M\gg T$ where these effects are negligible. The final baryon asymmetry is then proportional to $Y_{B-L}$ through the sphaleron conversion factor and inversely proportional to the entropy dilution factor
\begin{equation}
    Y_B=\frac{28}{79}\ \frac{Y_{B-L}}{\Delta}\ .
\end{equation}
The $B-L$ breaking scalar and the associated gauge boson $Z'$, do not play a significant role in leptogenesis in this regime, unless their own decays occur at very late times. This intriguing possibility is left for future investigation.
We now solve this set of Boltzmann equations for two regimes of leptogenesis: high-scale, where we assume hierarchical right-handed neutrino masses and low-scale, where we allow the masses to be fine-tuned whilst crucially avoiding the resonant regime.

\subsection{High-Scale Leptogenesis}
\label{sec:High_scale_lepto}
\begin{figure}[H]
    \centering
    \includegraphics[width=1\linewidth]{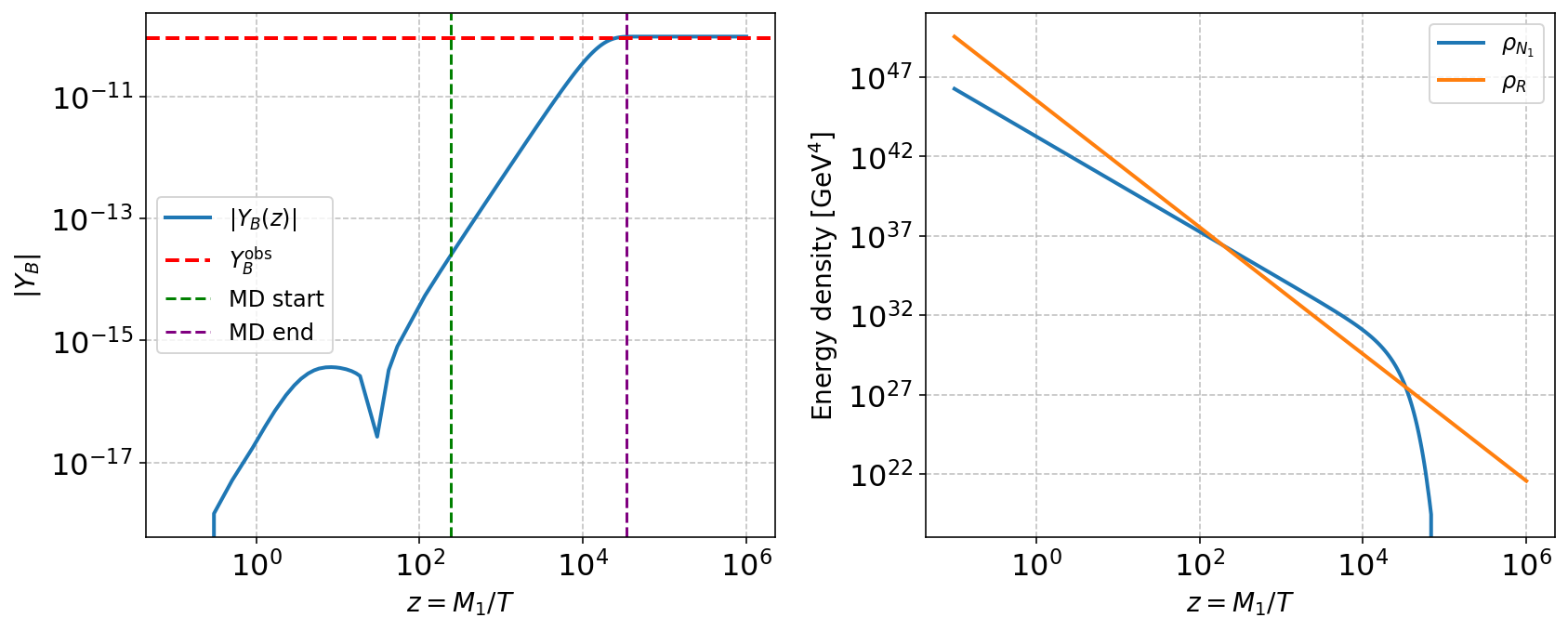}
    \caption{\it Benchmark for successful leptogenesis showing the evolution of the baryon asymmetry together with the right-handed-neutrino and radiation energy densities through a period of early matter domination. The vertical dashed lines indicate the onset and end of early matter domination. Parameters: $M_1=10^{11}\rm\ GeV$, $\tilde m_1=10^{-11}\rm\ eV$, and $\epsilon=\epsilon^{\rm DI}/5$. Note the kink around $z=10^2$ is due to a change in sign.}
    \label{fig:Highscale Benchmark 2}
\end{figure}

\begin{figure}[H]
    \centering
    \includegraphics[width=0.49\linewidth]{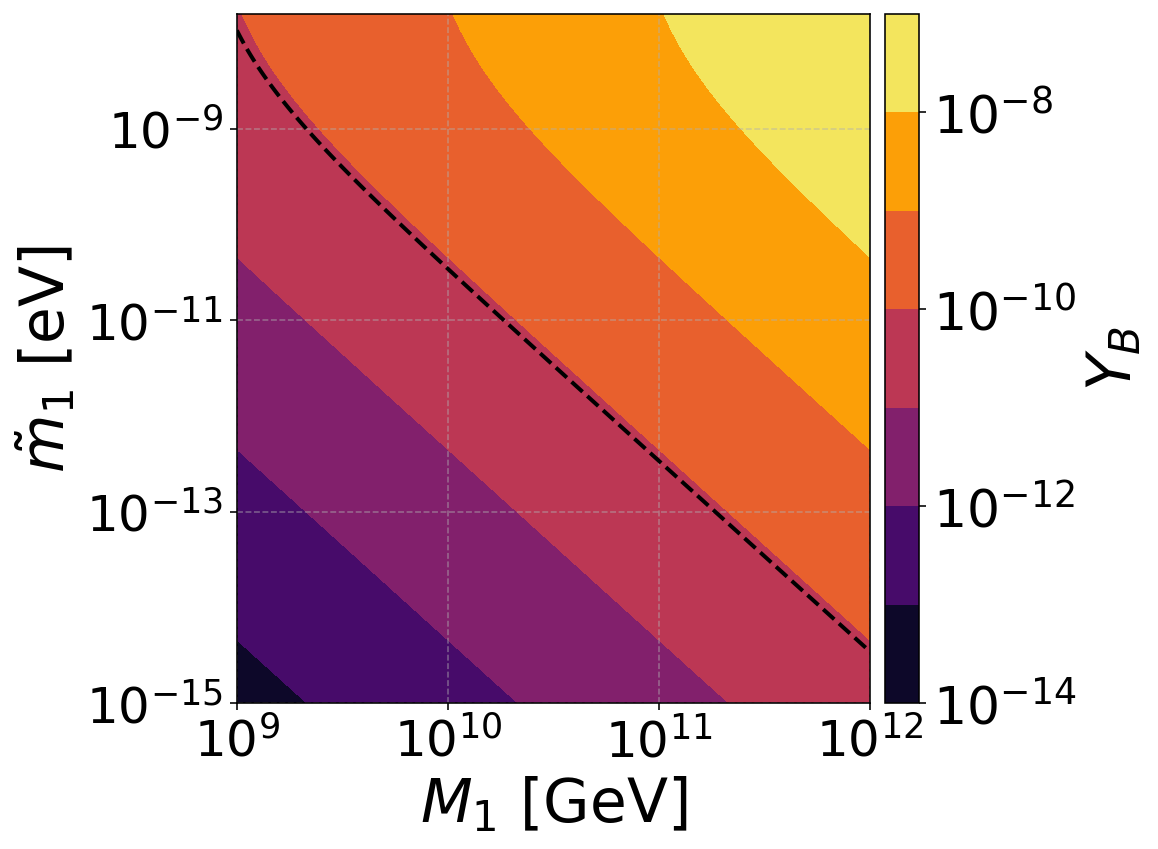}
    \includegraphics[width=0.49\linewidth]{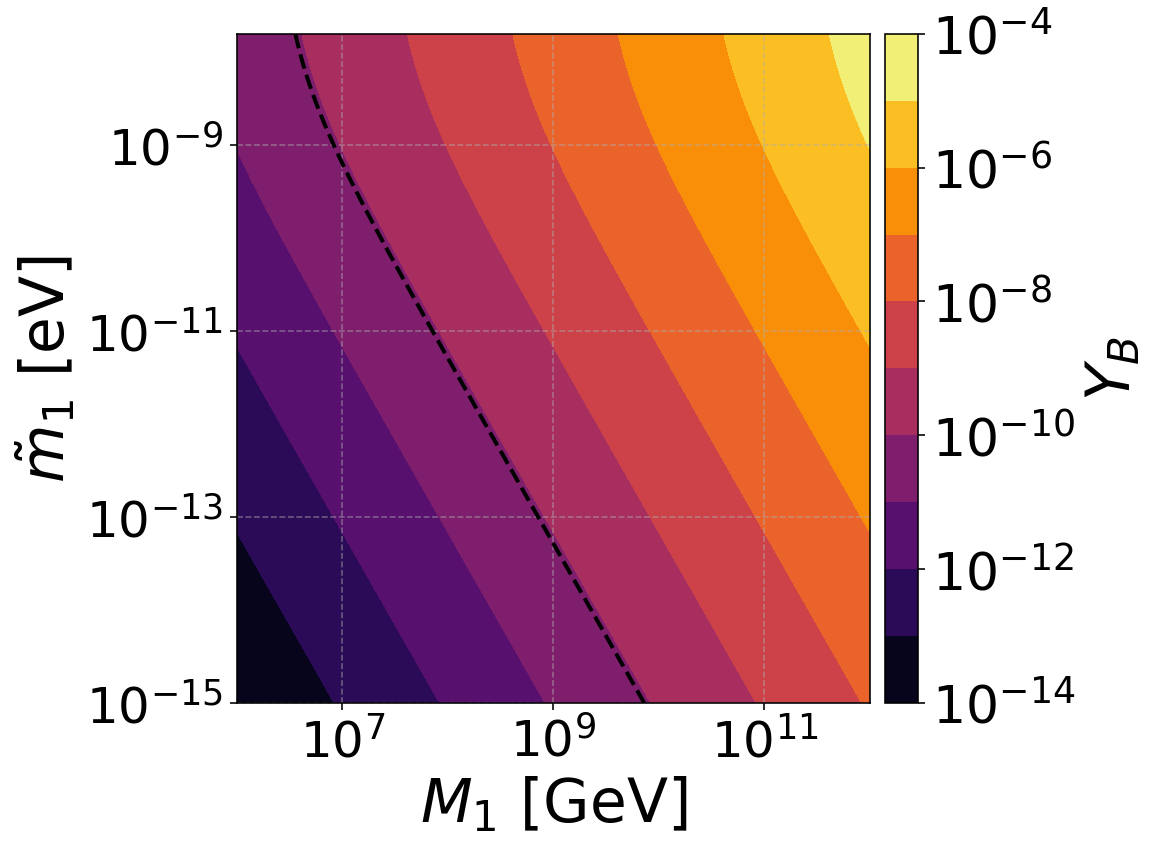}
    \caption{\it Parameter scans for successful leptogenesis with a period of early matter domination. The \textbf{left panel} corresponds to a thermal initial abundance, $Y_N^i=3.9\times 10^{-3}$, while the \textbf{right panel} assumes a non-thermal initial abundance, $Y_N^i=1$. The black dashed line denotes the observed baryon asymmetry; parameter space above this line yields successful leptogenesis.}
    \label{fig:Highscale}
\end{figure}

Assuming all the right-handed neutrinos decay before electroweak symmetry breaking, the calculation of the baryon asymmetry reduces to a simple relation involving the CP asymmetry parameter, initial abundance and entropy dilution factor and can be expressed as,
\begin{equation}
    Y_B = \frac{28}{79}\ \epsilon\ Y_{N}^{\text{init}}\  \frac{1}{\Delta}\ .
\end{equation}
To find the minimum right-handed neutrino mass for successful leptogenesis with a period of early matter domination, we wish to minimise the entropy dilution factor. Using Eq. \ref{eq:Delta}, this will occur when decays begin just when early matter domination begins so $\rho_R(t_{\rm dec})\approx\rho_N(t_{\rm dec})$.
\begin{equation}
    \text{Max}(\tfrac{1}{\Delta})=2^{-\frac{3}{4}}\approx 0.6
\end{equation}
Assuming a hierarchical mass spectrum of the right-handed neutrinos \( M_1 \ll M_2, M_3 \), and that the dominant lepton asymmetry arises from the decays of \( N_1 \), the CP asymmetry reduces to \cite{Fong_2012}
\begin{equation}
    \epsilon_1 = \frac{3}{16\pi} \ \frac{1}{(y^\dagger y)_{11}} \sum_{j \neq 1} \operatorname{Im} \left[ (y^\dagger y)_{j1}^2 \right] \ \frac{M_1}{M_j}
\end{equation}
subject to our condition for early matter domination Eq. \ref{eq: numeric mtilde}
\begin{equation}
    \tilde{m_1} < 1.1 \times 10^{-8}\, \mathrm{eV}\ .
\end{equation}
We performed a scan of $\tilde m_1$ inputs using a maximising function and found that even with the constraint, the maximum $\epsilon$ was the Davidson-Ibarra bound \cite{Davidson:2002qv} proportional to the mass of the lightest right-handed neutrino and the heaviest active neutrino,
\begin{equation}
    \epsilon^{\text{DI}} = \frac{3 M_1 m_3}{16\pi v_H^2}\ .
\end{equation}
With the maximum CP asymmetry parameter and minimum entropy injection, we find the minimum allowed $M_1$ is then
\begin{equation}
    M_1 > \frac{79}{28}\ \frac{16\pi v^2}{3m_3}\ \frac{Y_B}{Y_N^i}\ \frac{1}{\text{Min}(\Delta)}
\end{equation}
Taking hierarchical active neutrino masses, with the heaviest mass 
\( m_3 \simeq 0.05\,\mathrm{eV} \), and thermal initial abundance \(Y_{N_1}^{\rm init}\simeq 3.9\times 10^{-3}\), we obtain the lower bound of
\begin{equation}
    M_1 > 1.1 \times 10^9\ \mathrm{GeV}
\end{equation}
for the mass of the lightest right-handed neutrino. If instead we consider non-thermal production of right-handed neutrinos in generality, the bound becomes
\begin{equation}
    M_1>\frac{4.29 \times 10^6}{Y_N^i}\text{GeV}\ .
\end{equation}
Finally, the minimum effective neutrino mass allowed for successful thermal leptogenesis was calculated from these numerical solutions to be 
\begin{equation}
    \tilde{m_1} > 3.15 \times 10^{-15}\,\mathrm{eV}.
\end{equation}
We present a benchmark in Figure \ref{fig:Highscale Benchmark 2} for successful leptogenesis in this regime, showing the evolution of the baryon asymmetry through an intermediate period of early matter domination. The start and end of early matter domination are marked, providing testable scales that could be probed by future gravitational-wave experiments.
A complete scan of the $(M_1,\tilde{m}_1)$ parameter space, assuming the maximum CP asymmetry and a hierarchical mass spectrum, is shown in figure \ref{fig:Highscale} for both a thermal initial abundance and for a large non-thermal initial abundance with a period early matter domination. The scan shows that for all initial abundances, the baryon asymmetry increases with right-handed neutrino and effective neutrino mass. This is due to $\epsilon$ scaling with $M_1$ and the entropy dilution scaling inversely with $\tilde m$. A larger initial abundance broadens the parameter space in which leptogenesis can be successful, allowing for non-thermal production to lower the bound on the right-handed neutrino mass as well as the bound on the effective neutrino mass. The black dashed line indicates the observed baryon asymmetry; leptogenesis is viable in the region of parameter space above this line.
If we parametrise the true CP asymmetry parameter as $\epsilon=c\ \epsilon^{\rm DI},$ where $c\leq 1$ is a constant, then the mass for successful leptogenesis in the regime is fixed by the effective neutrino mass,
\begin{equation}
M_1 \;=\; \frac{6.39 \times 10^{8}}{c}\,
\left( 1 + 3.726 \times 10^{-6}\,\left(\frac{\tilde m}{\rm eV}\right)^{-0.67} \right)^{3/4}
\;\; \text{GeV}.
\end{equation}
In section \ref{sec:GW_test_lepto} we will show how this can be tested with gravitational wave spectral shapes. 
\subsection{Low-Scale Leptogenesis}
In the quasi-degenerate regime where $M_i \simeq M_j$, the loop function $f(x)$ strongly enhances the CP asymmetry. 
Expanding the form of $\epsilon$ in small $\delta M/M$ gives
\begin{equation}
\epsilon_i \;\simeq\; \frac{1}{16\pi}\,\frac{M_i}{\delta M}\,
\frac{Im \big[(y^\dagger y)_{ij}^2\big]}{(y^\dagger y)_{ii}},\ \leq \frac{1}{16\pi}\,\frac{M_i}{\delta M}\,(y^\dagger y)_{jj}.
\end{equation}
where in the last step we have bounded the numerator with the Cauchy-Schwartz inequality. $Im$ denotes imaginary component. As discussed in Sec.~\ref{sec: CP asymmetry}, we remain in the non-resonant regime to avoid the need for a regulator and the associated theoretical ambiguities of the resonant case, where the CP asymmetry is highly sensitive to the mass splitting and the chosen regularisation scheme. To remain in the non-resonant regime, the mass splitting must greatly exceed both decay widths \cite{Pilaftsis_2004, Moffat_2018},
\begin{equation}
\delta M > 100\,\Gamma_1,
\quad
\delta M > 100\,\Gamma_2, \quad \Gamma_i = \frac{(y^\dagger y)_{ii}}{8\pi}\,M_i\ .
\end{equation}
Evaluating $\epsilon_1$, the condition from $\Gamma_1$ yields an upper limit on the CP-violating parameter, expressed in terms of the ratio of effective neutrino masses,
\begin{equation}
|\epsilon_1| \;\lesssim\; \frac{1}{200}\,\frac{(y^\dagger y)_{22}}{(y^\dagger y)_{11}}\,\frac{M_2}{M_1}\simeq\frac{1}{200}\frac{\tilde m_2}{\tilde m_1}\ .
\end{equation}
In contrast, the constraint from $\Gamma_2$ removes any dependence on masses or Yukawas,
\begin{equation}
|\epsilon_1| \;\lesssim\; \frac{1}{200}.
\end{equation}
By symmetry, the corresponding limits for decays of the heavier right-handed neutrino are
\begin{equation}
|\epsilon_2| \;\lesssim\; \min\!\left[\frac{1}{200},
\;\frac{1}{200}\,\frac{\tilde m_1}{\tilde m_2}\right].
\end{equation}
Since both non-resonance conditions must hold simultaneously, the true universal ceiling is
\begin{equation}
|\epsilon_i| \;\lesssim\; \frac{1}{200}. 
\end{equation}
Thus, outside the resonant regime, the CP asymmetry per decay cannot exceed the percent level, independently of right-handed neutrino masses and Yukawa couplings. If one chooses to impose the resonance condition $\delta M > b \Gamma_i$ more or less strictly (with $b=100$ our baseline) the corresponding bound on the CP asymmetry follows directly as $\epsilon<1/2b$.

\begin{table}[h!]
\centering
\renewcommand{\arraystretch}{1.4}
\setlength{\tabcolsep}{10pt}
\begin{tabular}{|c|c|c|}
\hline
Hierarchy & Bound on $\epsilon_1$ & Bound on $\epsilon_2$ \\
\hline
$\tilde m_1 < \tilde m_2$ 
& $\displaystyle |\epsilon_1| \lesssim \frac{1}{200}$ 
& $\displaystyle |\epsilon_2| \lesssim \frac{1}{200}\,\frac{\tilde m_1}{\tilde m_2}$ \\
\hline
$\tilde m_2 < \tilde m_1$ 
& $\displaystyle |\epsilon_1| \lesssim \frac{1}{200}\,\frac{\tilde m_2}{\tilde m_1}$ 
& $\displaystyle |\epsilon_2| \lesssim \frac{1}{200}$ \\
\hline
\end{tabular}

\captionsetup{justification=raggedright, singlelinecheck=false, font=it}
\caption{Analytic bounds on the CP asymmetries, $\epsilon_{1,2}$, in the quasi-degenerate regime imposing the non-resonant condition $\delta M > 100 \Gamma_i$.}
\label{tab:epsbounds}
\end{table}

\begin{figure}[H]
    \centering
    \includegraphics[width=0.7\linewidth]{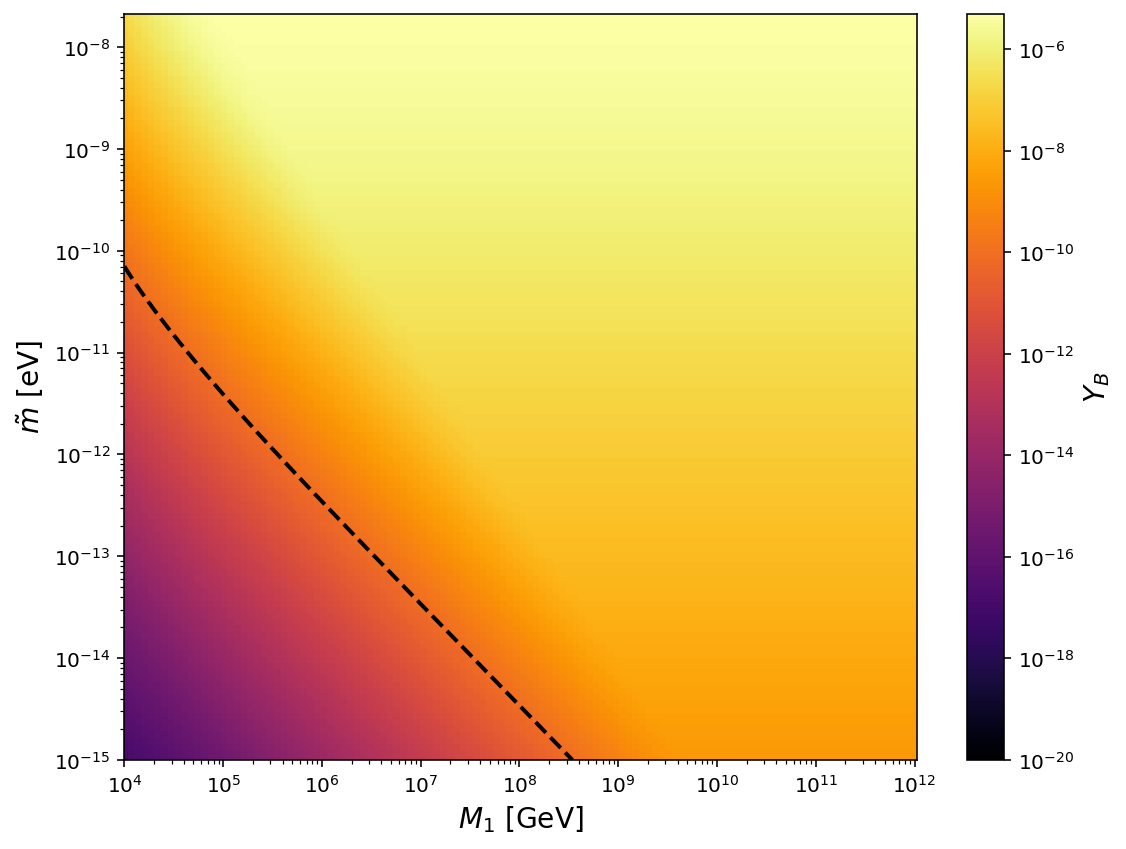}
    \caption{\it Parameter scan for near resonant leptogenesis. The black dashed line denotes the observed baryon asymmetry; parameter space above this line yields successful leptogenesis. In this regime, the baryon asymmetry is readily generated, leading to a much larger overlap with the region accessible to experiments compared to vanilla leptogenesis.}
    \label{fig:Lowscale scan}
\end{figure}

The state with the smaller effective mass $\tilde m$ saturates the universal ceiling $|\epsilon_i| \lesssim 1/200$, while the other is further suppressed by the ratio of effective neutrino masses. Since the earlier–decaying right-handed neutrino’s asymmetry is erased by washout processes involving the longer–lived species, only the decay of the right-handed neutrino with the smallest effective neutrino mass is relevant. In this regime, the CP asymmetry follows the universal bound $\epsilon<1/200$. As we are dealing with small right-handed neutrino masses, we cannot assume all the right-handed neutrinos have decayed by electroweak symmetry breaking; instead, we must solve the full Boltzmann equations with this bound on $\epsilon$ numerically up to the electroweak symmetry breaking temperature $T_{\rm EW}=130\rm\ GeV$ \cite{Ulysses2}. 
We performed a scan and show the parameter space in figure \ref{fig:Lowscale scan}. As in vanilla leptogenesis, the baryon asymmetry grows with both the right-handed-neutrino mass and the effective neutrino mass. Unlike the vanilla case, however, successful leptogenesis is realised over a much larger region of parameter space in the near-resonant case. This renders near-resonant leptogenesis far more testable than the vanilla scenario through gravitational wave observations.
In our analysis, we restrict the scan to $M_1 \gtrsim 10^{4}\,\text{GeV}$. This is because in this low-mass regime, the dynamics transition away from standard vanilla leptogenesis and approach the Akhmedov-Rubakov-Smirnov (ARS) mechanism \cite{Akhmedov_1998, Drewes_2018, Klaric_2021}, 
where lepton asymmetry is generated by oscillations of nearly degenerate right-handed neutrinos. Since our focus is on conventional leptogenesis rather than ARS leptogenesis, we conservatively impose the cut-off at $M_1 = 10^{4}\,\text{GeV}$.

\subsection{Primordial Gravitational Wave Tests of Leptogenesis} \label{sec:GW_test_lepto}
\begin{figure}[H]
\centering
\includegraphics[width=1\linewidth]{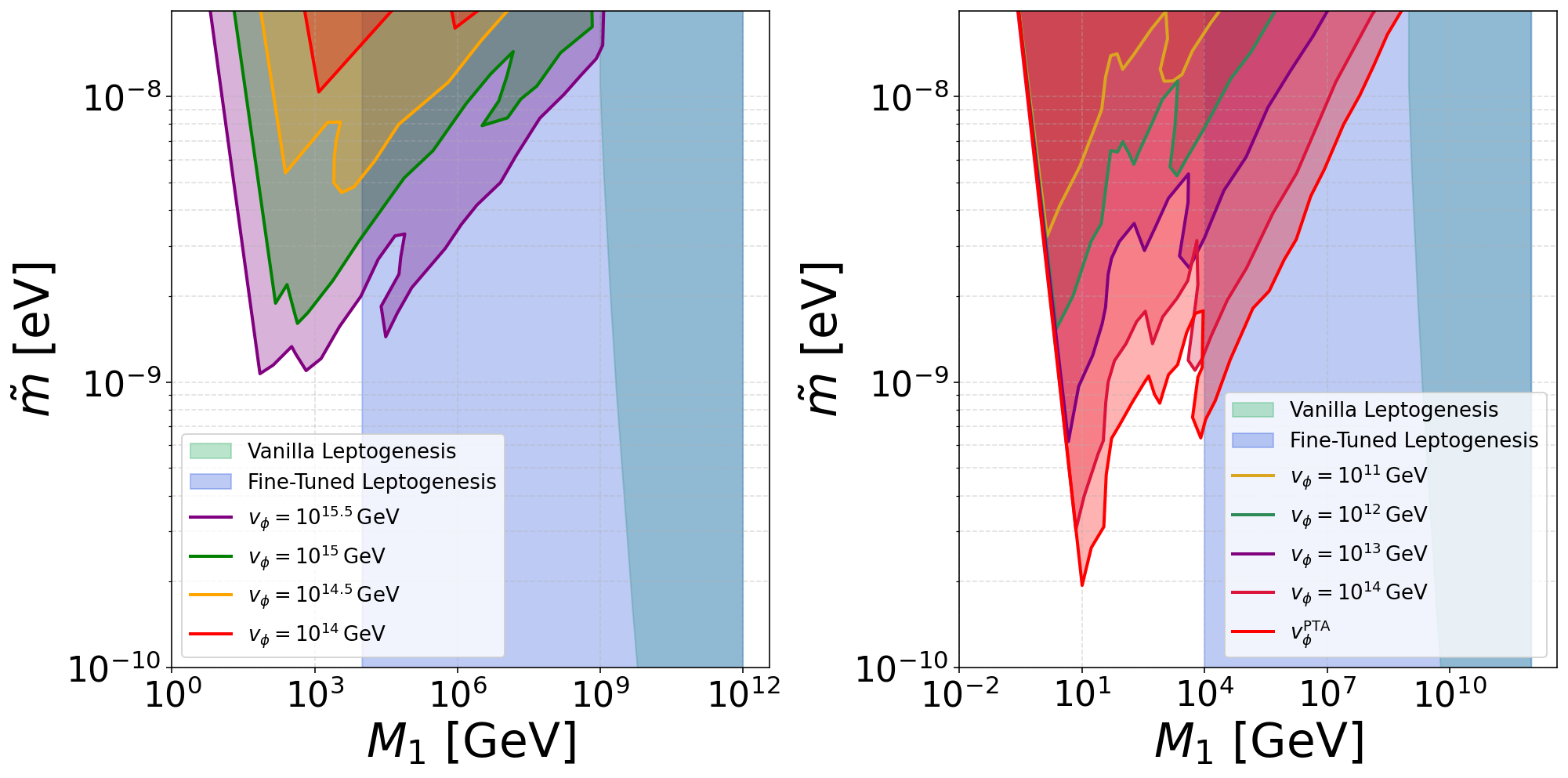} 
\caption{\it Detectable regions of the mass parameter space for leptogenesis with an intermediate period of early matter domination. The detectable kink in the gravitational-wave background arises from cosmic strings formed by the breaking of a global (\textbf{left panel}) or local (\textbf{right panel}) $U(1)_{B-L}$ symmetry. In both cases, the outer edges of the hierarchical regime, corresponding to the Davidson Ibarra bound, are just within reach, while a substantial fraction of the near-resonant parameter space is also detectable.}
    \label{fig:lepto experiment mass}
\end{figure}

Matching the conditions for successful leptogenesis in both the vanilla and resonant regimes with the experimental bounds on right-handed neutrino detectability defines the combined viable parameter space. The overlap of these detectable regions with the parameter space for successful leptogenesis, derived in section \ref{sec:cosmicstrings}, is illustrated in figures \ref{fig:lepto experiment mass}. 

Figure \ref{fig:lepto experiment mass} shows that gravitational waves can probe only the outer edges of the hierarchical leptogenesis parameter space, while in the near-resonant regime they are sensitive to a much broader region of viable models. There are two interesting observables from the GWB spectral shape, $f_{\rm dom}$ and $f_{\rm brk}$. The first of these frequencies of interest, $f_{\rm dom}$, is characteristic of the onset of early matter domination and is therefore determined by the right-handed neutrino mass $M_1$. In the hierarchical regime, large $M_1$ values $M_1 > \mathcal{O}(10^{10})\rm\ GeV$ lead to very high $T_{\text{dom}}$, which shifts $f_{\text{dom}}$ beyond the reach of next-generation detectors. Consequently, only the edges of the hierarchical parameter space, with smaller $M_1$, produces a detectable feature. In the near-resonant regime, successful leptogenesis can occur for much smaller right-handed-neutrino masses, moving $T_{\rm dom}$ to lower values and placing $f_{\text{dom}}$ well within the sensitivity range of interferometers. The near-resonant scenario, therefore, provides a broad and accessible observational target.

The other frequency, $f_{\text{brk}}$, is characteristic of the end of early matter domination at $T_{\text{end}}$, which scales with $M_1$ and depends on the effective neutrino mass $\tilde m_1$ through the decay rate of the right-handed neutrinos.  Shorter periods of early matter domination are more favourable for detection, since they produce higher break frequencies within the sensitivity range of future interferometers. Shorter periods, corresponding to $\tilde m_1$ slightly below but close to $10^{-8}\ \mathrm{eV}$, yield a narrow and well-defined feature that falls within the sensitivity bands of upcoming detectors. These frequencies, therefore, encode both the scale and the duration of sterile-neutrino domination. Importantly, such regions cannot be accessed through collider or low-energy neutrino experiments; gravitational-wave observations provide the only direct probe of this otherwise hidden sector of leptogenesis.

\medskip

\section{Gravitational Wave Tests of Dark Matter Formation}
\label{sec:dark matter}

 In addition to explaining neutrino masses and baryogenesis, the Type-I seesaw framework can be naturally extended to accommodate dark matter \cite{cirelli2024darkmatter}. We consider such extensions in this section with an
eye on how the possibility of an early period of matter domination affects the scenario. Cosmological observations firmly establish a non-baryonic component with relic density $\Omega_{\rm DM} h^{2} \simeq 0.12$ \cite{cirelli2024darkmatter}. While thermal freeze-out is the canonical paradigm \cite{cirelli2024darkmatter, Bertone_2005, Kolb:1990vq, Weinberg:2008zzc, Zeldovich:1965gev, Bernstein:1985th, Scherrer:1985zt, Gondolo:1990dk, Hisano_2007, Cirelli_2007}, non-thermal production via late decays provides a robust alternative that naturally arises in neutrino-mass models \cite{Falkowski_2011, Dutta_Banik_2021, Barman_2022}. In this work, the right-handed neutrino $N$ drives an early matter-dominated epoch and subsequently decays out of equilibrium into both the Standard Model and a dark sector. Concretely, we augment the $B-L$ Type I seesaw Lagrangian, Eq. \ref{eq: TypeI Lagrangian}, by a Yukawa interaction
\begin{equation}
    \mathcal{L} \supset y_{\rm DM}\, \bar{N}\,\chi\,\eta + \text{h.c.},
\end{equation}
where $\chi$ is the stable dark matter candidate that is a fermion and $\eta$ a scalar, and $y_{\rm DM}$ is the interaction strength between $N$ and the DM sector. It is straightforward to give $\eta$ decay modes meaning only $\chi$ is a dark matter candidate. The simplest way is to have a mixed quartic portal coupling with the Higgs and give $\eta$ a vacuum expectation value at some stage in its history. These details we leave unspecified to focus on the phenomenology of the right-handed neutrinos and the dark matter candidate. The corresponding mass terms are left implicit, allowing the dark matter candidate to be either a scalar or a fermion, with its mass possibly arising from the $B-L$ symmetry breaking or from an independent mechanism.\footnote{For instance, in the local case, $M_{\rm DM} \propto v_{\rm B-L}$}. This distinction is inconsequential for the dark matter phenomenology discussed here. The same $N$ also decays through its seesaw couplings into Higgs and Lepton pairs. If $N$ dominates the energy density prior to its decay, the ensuing entropy injection becomes integral to the relic prediction: the dark matter yield is set by the pre–decay abundance of $N$ and the branching fraction into $\chi$, diluted by the entropy released at decay. This ties the relic density to a small set of parameters $(y_{\rm DM}, M, \tilde m)$ and the expansion history, enabling analytic control over $T_{\rm dec}$ and the dilution factor $\Delta$.

\subsection{Lower bound on Dark Matter Mass}
\begin{figure}[H]
\centering
\includegraphics[width=0.49\linewidth]{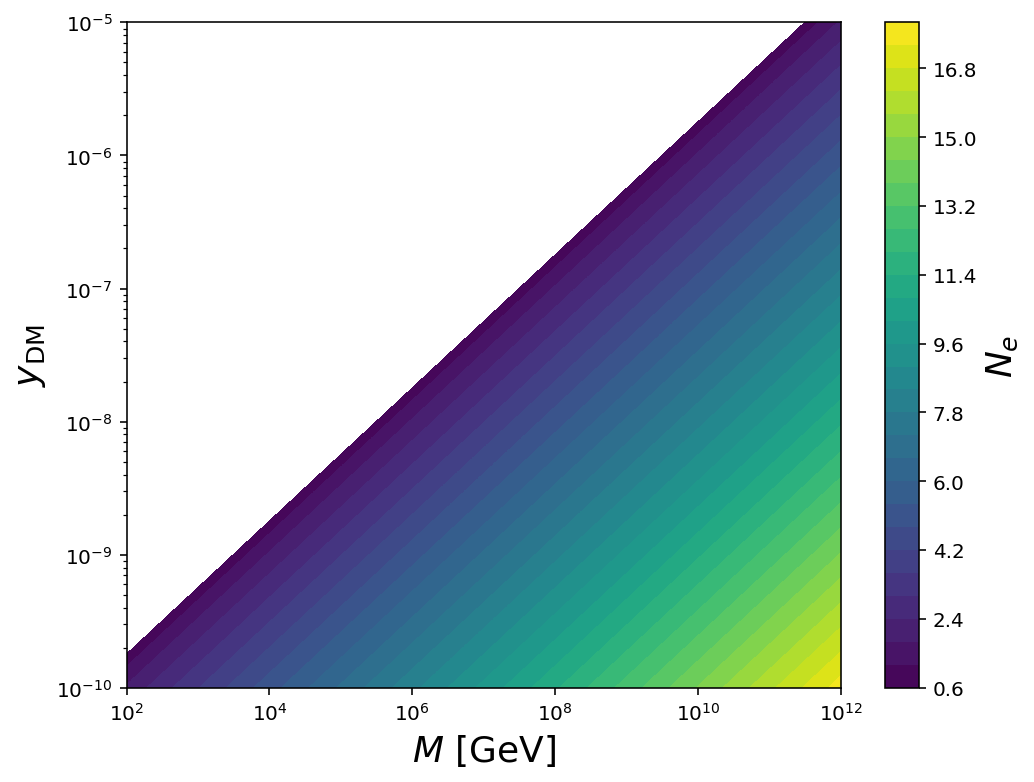} 
\includegraphics[width=0.49\linewidth]{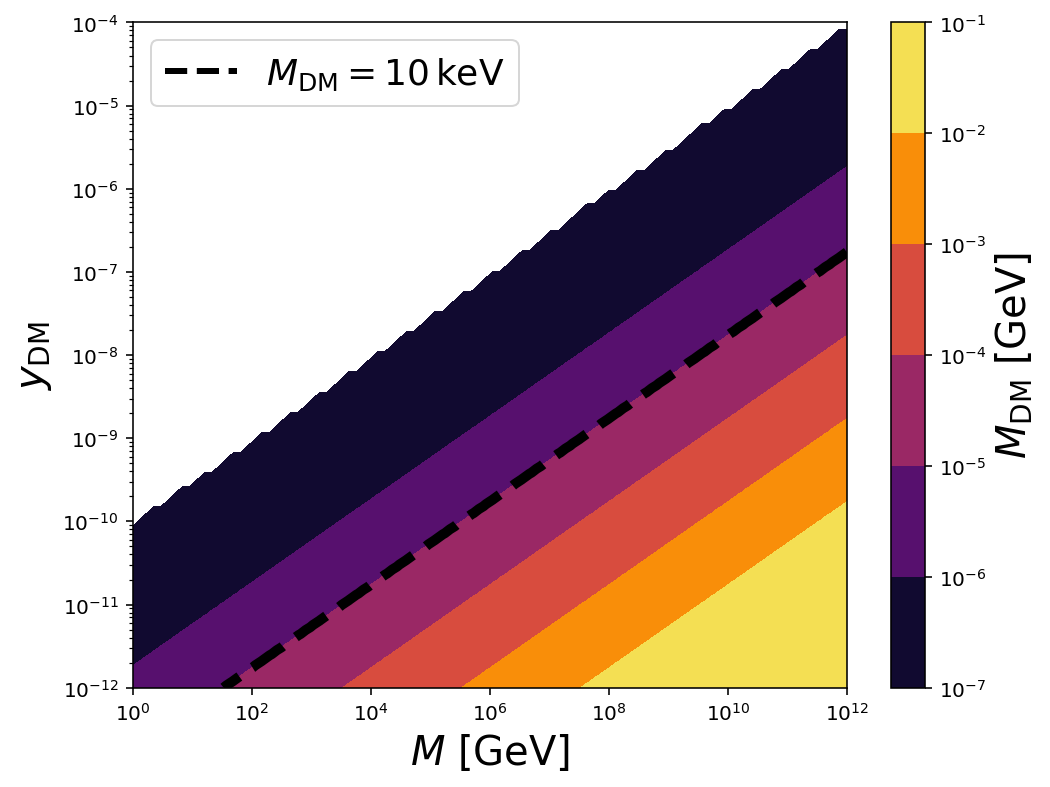}  
\caption{\it \textbf{Left:} $N_e$ as a function of $y_{DM}$ and $M$. \textbf{Right:} Dark matter mass assuming a branching ratio $\approx 1$. A longer period of early matter domination increases entropy dilution, which suppresses the dark matter yield and thus requires a larger dark matter mass. The black dashed line indicates the approximate warm dark matter mass bound. For the assumed branching ratio, the viability of the dark matter candidate requires a period of early matter domination. In both plots, only parameter points that feature a period of early matter domination are shown.}
    \label{fig:DM efolds}
\end{figure}

\noindent If we assume that after dark matter production, there is standard cosmology, the observed dark matter yield is inversely proportional to the dark matter mass ~\cite{cirelli2024darkmatter},
\begin{equation}
    Y_{\mathrm{DM}} \;\approx\; 4.37 \times 10^{-10}\;
    \left(\frac{\Omega_{\mathrm{DM}} h^{2}}{0.12}\right)\!
    \left(\frac{\mathrm{GeV}}{M_{\mathrm{DM}}}\right)\ .
\end{equation}
Here $\Omega_{\mathrm{DM}}$ is the present abundance and $M_{\rm DM}$ is the dark matter mass. The inverse scaling with $M_{\mathrm{DM}}$ implies that any dilution of the yield, in our case due to entropy injection during an early period of matter domination, must be compensated by a correspondingly larger dark matter mass in order to reproduce the observed relic abundance.
If we, for now, neglect decays into the Standard Model particles, assume that all right-handed neutrinos decay into dark matter, we can obtain a lower bound on the dark matter mass under the condition of minimal entropy injection from the matter-domination era, while the upper bound comes from the necessity for decays to be allowed. For thermal production \footnote{In this section, we restrict our analysis to a thermal initial abundance of right-handed neutrinos. } this lower bound is weaker than Lyman-$\alpha$ bounds on warm dark matter species \cite{Dvorkin_2021, Decant_2022, Berbig:2023yyy}, the range of allowed masses of dark matter is then
\begin{equation}
    M>M_{\rm DM}>\mathcal{O}(10\ \text{KeV})\ .
\end{equation}
This demonstrates the viability of dark matter to be included in our model. We consider the right-handed neutrino mass to be much larger than those of $\chi$ and $\eta$. The production rate of $\chi$ is then proportional to the mass of the right-handed neutrino and the squared magnitude of the Yukawa coupling,
\begin{equation}
    \Gamma_{\chi} = \frac{|y_{\rm DM}|^2 M}{8\pi}\ ,
\end{equation}
and early matter domination condition following the logic of section \ref{sec:domination} is a condition on the Yukawa coupling and the mass,
\begin{equation}
    |y_{\rm DM}|\lesssim 9.5\times 10^{-11}\ \sqrt{\frac{M}{\text{GeV}}}
\end{equation}
These bounds are easily satisfied for even a large right-handed neutrino mass, showing that we can achieve successful dark matter and a period of early matter domination for high-scale masses. We performed a systematic scan over the parameter space $(y_{\rm DM}, M)$ in order to quantify the impact of the dark-sector Yukawa coupling and the right-handed neutrino mass on the duration of the matter-dominated epoch and the entropy injection. The results found that both $N_e$ and, as such, the entropy dilution increased with right-handed neutrino mass and decreased with Yukawa coupling; the results for $N_e$ are shown in the left panel of Fig.~\ref{fig:DM efolds}.
By matching the results to the observed dark matter abundance, we determine the parametric dependence of the dark matter mass on the right-handed neutrino parameters. As shown in the right panel of Fig.~\ref{fig:DM efolds}, the mass of the dark matter candidate grows with increasing right-handed neutrino mass and decreases with larger Yukawas. This behaviour reflects the fact that a longer period of early matter domination enhances entropy injection, thereby reducing the dark matter yield and necessitating a heavier dark matter mass. The plot assumes a branching ratio of unity. In this case, a viable dark matter candidate only emerges if the universe undergoes a period of early matter domination and the subsequent entropy dilution, this is shown in the $M_{\rm DM}>\mathcal{O}(10 )$ keV bound in the right panel of Fig ~\ref{fig:DM efolds}. More generally, if the branching ratio into dark matter is below $1\%$, this requirement is lifted; but once it exceeds $1\%$, early matter domination and the subsequent entropy dilution become essential for dark matter to remain viable.

\subsection{Lower bound on Asymmetric Dark Matter Mass}

One of the most striking features of the Universe is the near coincidence between the baryonic and dark matter energy densities, $\Omega_{\mathrm{DM}} / \Omega_B \simeq 5$. Despite their apparently distinct origins, this numerical proximity hints at a common mechanism that simultaneously generates both components of cosmic matter. Asymmetric dark matter (ADM) models \cite{Falkowski_2011, Dutta_Banik_2021, Barman_2022} offer a natural explanation: the dark matter relic abundance arises from an initial matter–antimatter asymmetry rather than thermal freeze-out, mirroring the generation of the baryon asymmetry through leptogenesis. In such frameworks, the CP-violating decays of heavy right-handed neutrinos directly produce lepton asymmetries in both the visible and dark sectors. The lepton asymmetry in the Standard Model is subsequently converted into baryon number via sphaleron processes, while the corresponding dark asymmetry determines the relic density of the dark sector. This mechanism elegantly unifies the explanations for neutrino masses, the baryon asymmetry of the Universe, and the dark matter abundance within a single extension of the Standard Model, suggesting that the cosmic composition of matter ultimately traces back to the dynamics of sterile neutrino decays.\\
To implement this model, we must couple the dark matter particle to at least two right-handed neutrinos, which we denote as $N_1,\ N_2$ with Masses $M_1,\ M_2$ respectively,
\begin{equation}
    \mathcal{L}\subset -\mathcal{Y}_i N_i \chi \eta
\end{equation}
Following \cite{Falkowski_2011}, the dark matter abundance will depend on the dark matter CP asymmetry, efficiency, as well as initial abundance and entropy dilution
\begin{equation}
    Y_{DM}=\frac{\epsilon_{\chi}\eta_\chi Y_N^i}{\Delta}\ .
\end{equation}
We are assuming that decays are very late and mainly into dark matter, not to $H,L$, so we can take $\eta_\chi\approx 1$. The dark matter CP asymmetry parameter, $\epsilon_{\chi}$, is defined in an analogous way to the leptogenesis CP asymmetry parameter in terms of decay rates.
\begin{equation}
    \epsilon_\chi = \frac{\Gamma(N_1 \to \chi \eta) - \Gamma(N_1 \to \bar{\chi} \eta^\dagger)}{\Gamma_{N_1}}.
\end{equation}
For hierarchical right-handed neutrino masses, the parameter \(\epsilon_{\chi}\) is dependent on these masses and the Yukawa couplings \cite{Falkowski_2011}, and can be expressed as,
\begin{equation}
    \epsilon_{\chi}\simeq \frac{M_1 \mathcal{Y}_2^2}{16\pi M_2}.
\end{equation}
Since we are solving for the lower bound, we once again take the minimum entropy injection and arrive at 
\begin{equation}
    M_{\rm DM} \;\gtrsim\;
    \frac{3.66\times 10^{-8}}{Y_N^i}\,
    \frac{M_2}{\mathcal{Y}_2^2 M_1}\ \mathrm{GeV},
\end{equation}
which shows it is straightforward and simple for asymmetric dark matter to be included. As an example, taking $M_2=10$, $M_1=10^{10}\rm \ GeV$ and thermal initial abundance, the condition on the dark matter mass is
\begin{equation}
    M_{\rm DM}>\frac{10^{-19}}{\mathcal{Y}_2^2}\ ,
\end{equation}
which imposing the condition $M_{\rm DM}<M_1$ has the bound for successful early matter domination and leptogenesis and asymmetric dark matter to be, $\mathcal{Y}_2\gtrsim 10^{-14}$ showing these conditions are trivially satisfied. 
\subsection{Co-genesis of Baryon Asymmetry and Dark Matter Asymmetry}
In this section, we identify the parameter space that permits the simultaneous generation of the baryon asymmetry and the dark matter abundance with a period of early matter domination. The entire framework ultimately traces its origin to the \( U(1)_{B-L} \) symmetry. Its spontaneous breaking sets the scale for the right-handed neutrino masses, which govern both leptogenesis and dark matter production, and simultaneously generates the cosmic strings that source the stochastic gravitational wave background. The same mass scale determines when the right-handed neutrinos dominate the energy density, thereby modifying the expansion history and imprinting distinctive features on the gravitational wave spectrum. In this way, the right-handed neutrino masses, leptogenesis, dark matter, gravitational waves and the spectral shape modification are unified through their common origin in the \( U(1)_{B-L} \) breaking scale.
To consider this full framework and the co-genesis of baryon asymmetry and dark matter, we specify that the right-handed neutrino $N$ from the previous analysis is required to be the lightest right-handed neutrino $N_1$. To proceed with the analysis, it is convenient to introduce a new variable. To this end, we recall, Eq. \ref{eq: Decay rate}, that the decay rate of a right-handed neutrino $N$ into leptons is
\begin{equation}
    \Gamma_{N \to HL} \;=\; \frac{(y^\dagger y)_{ii}}{8\pi}\,M \,,
\end{equation}

and it is beneficial to define the effective neutrino mass
\begin{equation}
    \tilde m \;\equiv\; \frac{(y^\dagger y)_{ii}\,v_H^2}{M} \,,
\end{equation}
which controls both the right-handed neutrino decay width and its connection to the light neutrino masses in the seesaw mechanism and the condition for early matter domination. If the right-handed neutrino also couples to a dark matter candidate $\chi$ with a Yukawa coupling $y_{\rm DM}$, the additional decay channel $\Gamma_{N \to \chi}$ 
can be expressed in terms of an analogous parameter,
which we denote the effective dark matter mass parameter,
\begin{equation}
    \tilde m_\chi \;\equiv\; \frac{y_{\rm DM}^2\,v_H^2}{M} \,.
\end{equation}

\begin{figure}[H]
\centering
\includegraphics[width=1.02\linewidth]{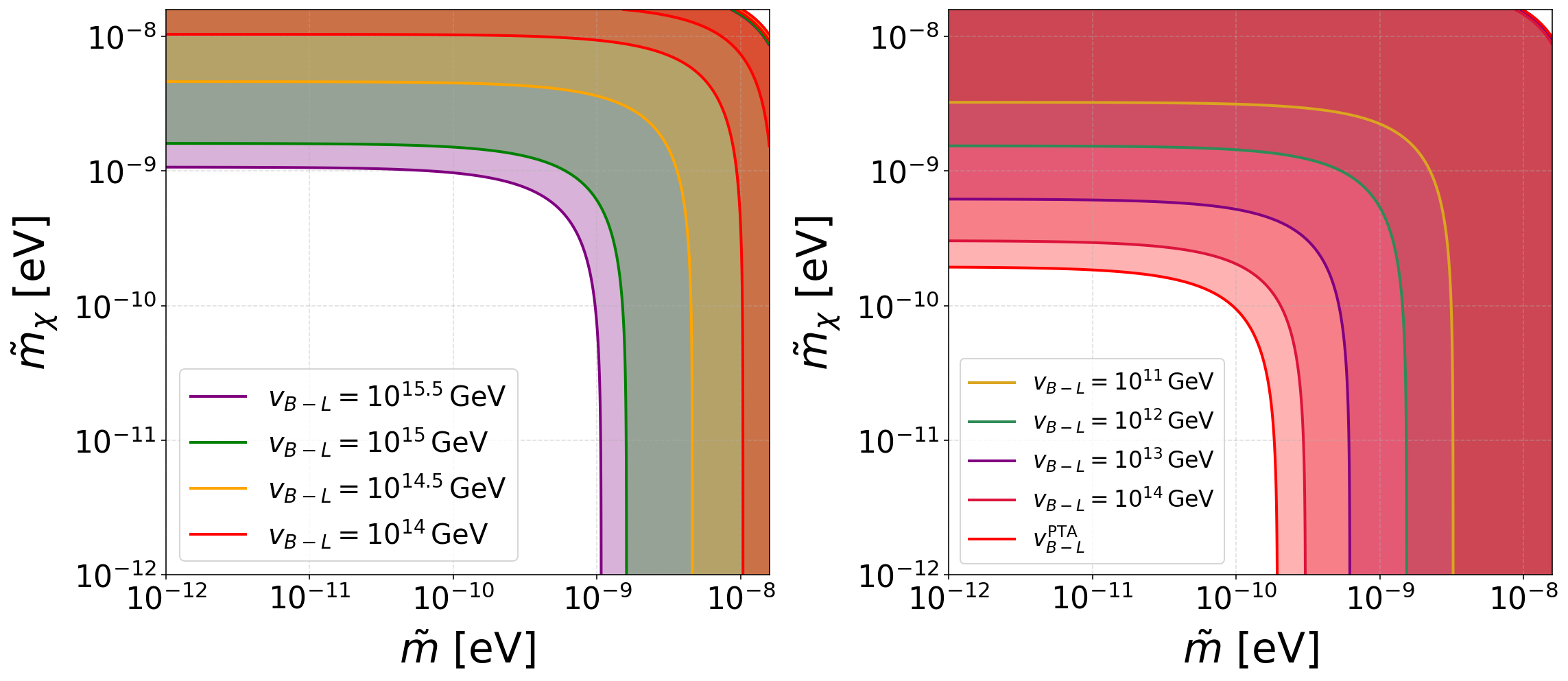} 
\caption{\it The detectable range with future gravitational wave experiments from global cosmic strings \textbf{(Left)} and local cosmic strings \textbf{(Right)}. The relevant parameter space is spanned by the Type-I effective neutrino mass, $\tilde{m}$, and the newly defined effective dark matter mass parameter, $\tilde{m}_{\chi} \approx \frac{y_{\text{DM}}^{2} v_H^{2}}{M}$. Global strings probe down to $\tilde m_{tot}\approx 10^{-9}$ eV, whereas local strings probe down to $\tilde m_{tot}=10^{-10}$ eV. }
    \label{fig:DM effective mass experiment}
\end{figure}

\begin{figure}[H]
\centering
\includegraphics[width=0.7\linewidth]{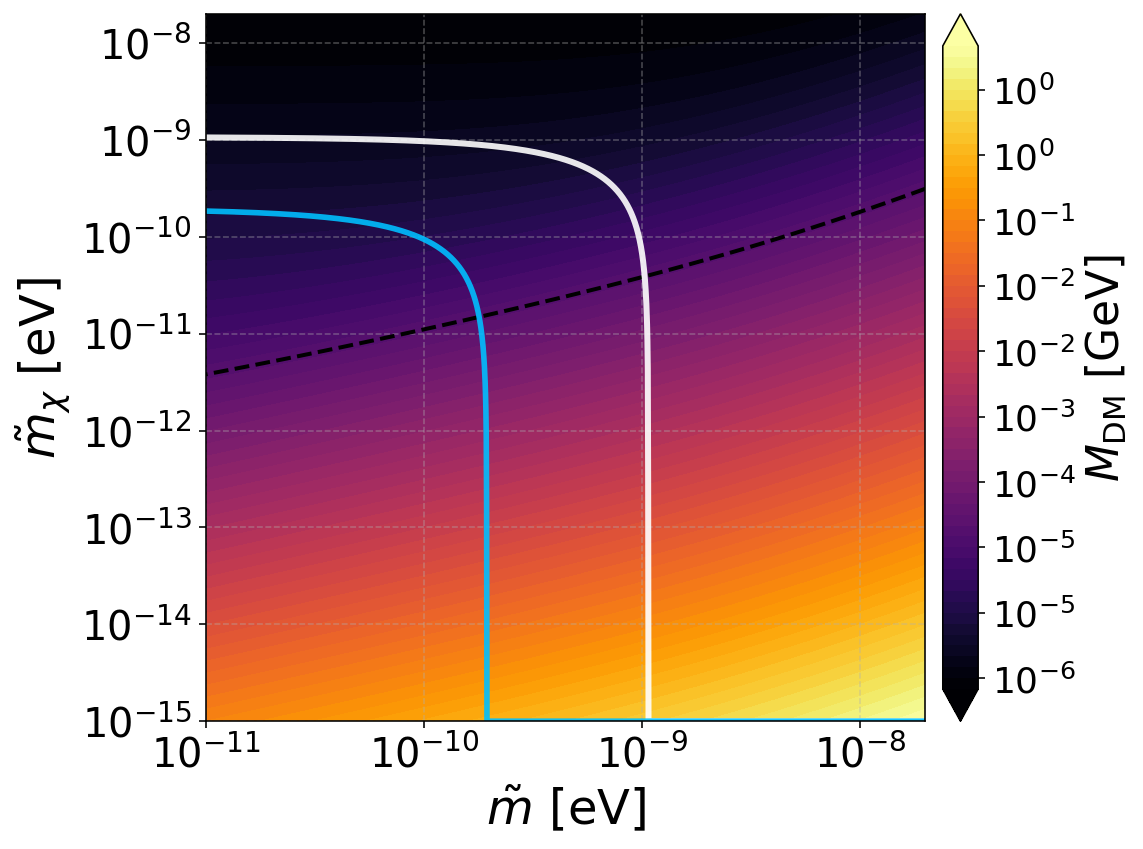} 
\caption{\it The mass of the dark matter candidate dependence on the effective mass parameters $\tilde m$ and $\tilde m_\chi=\frac{y_{\rm DM}^2v_H^2}{M}$. Warm dark matter is ruled out above the $M_{\rm DM}\approx 10 \rm $ keV dashed black line. The area to the right and above the blue and white lines indicates the regions that lead to detectable GW signals for local and global strings, respectively. This demonstrates the relative simplicity of dark matter production and how readily it can be tested through local and global $U(1)_{B-L}$ gravitational-wave backgrounds, see Fig. \ref{fig:DM effective mass experiment} for details. }
    \label{fig:DM effective mass scan}
\end{figure}

This parameter is introduced to reduce the original three-variable system, $(y_{\rm DM}, \tilde m, M)$, to a two-variable one, $(\tilde m_\chi, \tilde m)$, simplifying the analysis and making the experimentally testable parameter space more transparent. It also allows the relations derived in Section 2 to be directly applied in this context. We refer to it as an effective mass due to its dimensionality and its formal resemblance to the Type-I seesaw effective neutrino mass. However, this quantity is a theoretical construct introduced for convenience and is unrelated to the seesaw mechanism itself. The total decay width of the right-handed neutrino is then
\begin{equation}
    \Gamma^{\rm tot} \;=\; \frac{M^2}{8\pi v_H^2}\,\tilde m_{\rm tot}\,,\quad \tilde m_{\rm tot} \;\equiv\; \tilde m + \tilde m_\chi \,.
\end{equation}
This parameterisation then allows every result in section \ref{sec:domination} to be reproduced with the replacement $\tilde m\rightarrow \tilde m_{\rm tot}$. The condition for early matter domination becomes,
\begin{equation}
    \tilde m_{\rm tot}<1.1\times 10^{-8}\rm\ eV\ ,
\end{equation}
and the ending of early matter domination occurs at, 
\begin{equation}
    T_{\rm end}(M, \tilde{m}_{\rm tot}) \;\approx\; 
    4.5 \times 10^{-3} \, M \,
    \left( \frac{5.22 \times 10^{-8}\ \rm eV}{\tilde{m}_{\rm tot}} - 1 \right)^{-0.68}
\end{equation}
and the entropy dilution is given by,
\begin{equation}
    \Delta = \left( 1 + 3.726 \times 10^{-6} \,\left(\frac{\tilde m_{\rm tot}}{\rm eV}\right)^{-0.67} \right)^{3/4}\ .
\end{equation}
Importantly, this formulation allows for a straightforward determination of the gravitational-wave detection parameter space, shown in Fig.~\ref{fig:DM effective mass experiment}. The figure demonstrates that while global strings offer quite a limited probe, sensitive only to the boundary of the matter-dominated region, local cosmic strings can probe a much larger fraction of the parameter space. Thus, local strings offer a significantly stronger and more comprehensive test of the underlying $U(1)_{B-L}$ dynamics than their global counterparts.
The branching ratios are then obtained by taking the ratio of the partial widths to the total width:
\begin{equation}
    \text{Br}_{\chi} \;=\;  \text{Br}(N \to HL) \;=\; \frac{\Gamma_{N \to HL}}{\Gamma_N^{\rm tot}}
    \;=\; \frac{\tilde m}{\tilde m + \tilde m_\chi} \,,
\end{equation}
\begin{equation}
    \text{Br}_{HL} \;=\; \text{Br}(N \to \chi) \;=\; \frac{\Gamma_{N \to \chi}}{\Gamma_N^{\rm tot}}
    \;=\; \frac{\tilde m_\chi}{\tilde m + \tilde m_\chi} \,.
\end{equation}
The solutions for the dark matter abundance and the baryon asymmetry of the universe become analytic with this parameterisation,
\begin{equation}
    Y_{\rm DM}=\text{Br}_{\chi}\ Y_N^i\ \frac{1}{\Delta},\qquad Y_B =\frac{28}{79}\ \epsilon\ \text{Br}_{HL}\ Y_N^i\ \frac{1}{\Delta}
    \label{eq: DM and Lepto analytics}
\end{equation}
The dark matter mass, which has to be less than the right-handed neutrino mass, is determined uniquely by the two effective mass parameters, which we show in figure \ref{fig:DM effective mass scan}. 

This demonstrates how a smaller effective dark matter mass parameter and a larger effective neutrino mass lead to a larger dark matter mass. From the contour analysis and imposing the bound $M_{\rm DM}>\mathcal{O}(10\rm KeV)$ from Lyman-$\alpha$ analysis, we find that viable dark matter requires $\tilde m_\chi\lesssim 3\times 10^{-10}$.
This analysis can be straightforwardly extended to the case of a non-thermal initial abundance of right-handed neutrinos, where $Y_N^i$ in Eq.~\ref{eq: DM and Lepto analytics} is replaced by the corresponding non-thermal abundance specific to the model under consideration. An increased initial abundance enhances the resulting dark matter yield, thereby requiring a smaller dark matter mass to reproduce the observed relic density. The dark matter abundance can be expressed in two equivalent forms: one in terms of the effective masses and the initial abundance, and the other in terms of the dark matter mass, which can be written as follows
\begin{equation}
\begin{aligned}
    Y_{\rm DM}&\simeq4.37 \times 10^{-10}\;
    \left(\frac{\Omega_{\mathrm{DM}} h^{2}}{0.12}\right)\!
    \left(\frac{\mathrm{GeV}}{M_{\mathrm{DM}}}\right)\\
    &\simeq\frac{\tilde m_\chi}{\tilde m_{tot}}\ \left( 1 + 3.726 \times 10^{-6} \,\left(\frac{\tilde m_{\rm tot}}{\rm eV}\right)^{-0.67} \right)^{-3/4}\ Y_N^i\ .
\end{aligned}
\end{equation}
This shows that the dark matter mass is inversely proportional to the initial abundance of right-handed neutrinos, making it straightforward to accommodate non-thermal initial conditions.

\begin{figure}[H]
\centering
\includegraphics[width=1\linewidth]{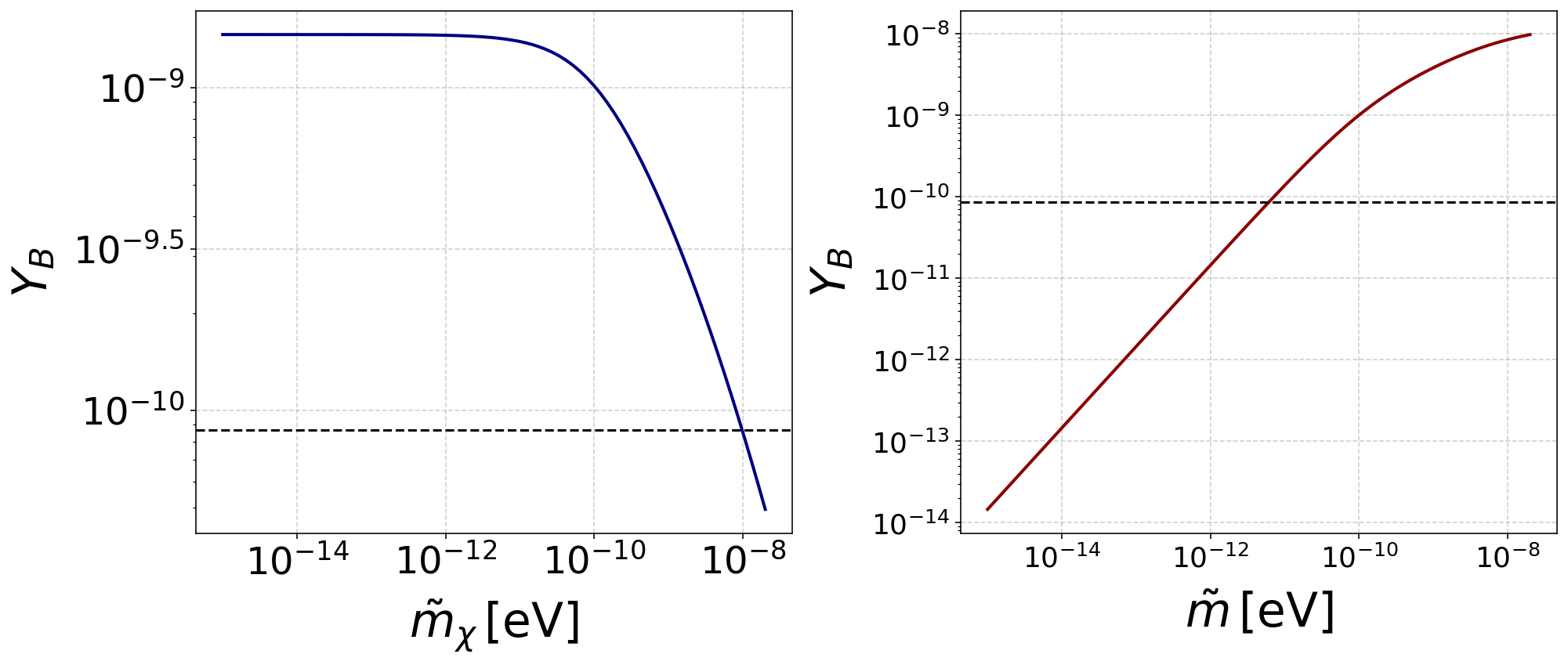} 
\caption{\it Demonstration of the dependence of the baryon asymmetry on the effective masses, $\tilde m$ and $\tilde m_\chi=\frac{y_{\rm DM}^2v_H^2}{M_1}$. We fix $M_{1}=10^{11}\,\text{GeV}$ and one effective mass parameter to $10^{-10}\,\text{eV}$, while varying the other. The left panel shows the dependence on $\tilde m_{\chi}$ (with $\tilde m=10^{-10}\,\text{eV}$ fixed), and the right panel shows the dependence on $\tilde m$ (with $\tilde m_{\chi}=10^{-10}\,\text{eV}$ fixed). In both cases, the observed asymmetry is indicated by the horizontal dashed line. Increasing \( \tilde{m} \) and decreasing \( \tilde{m}_\chi \) enhance the branching ratio of the Higgs--lepton decay channel, thereby leading to a larger baryon asymmetry.
}
    \label{fig:lepto effective mass scan}
\end{figure}

\begin{figure}[H]
\centering
\includegraphics[width=\linewidth]{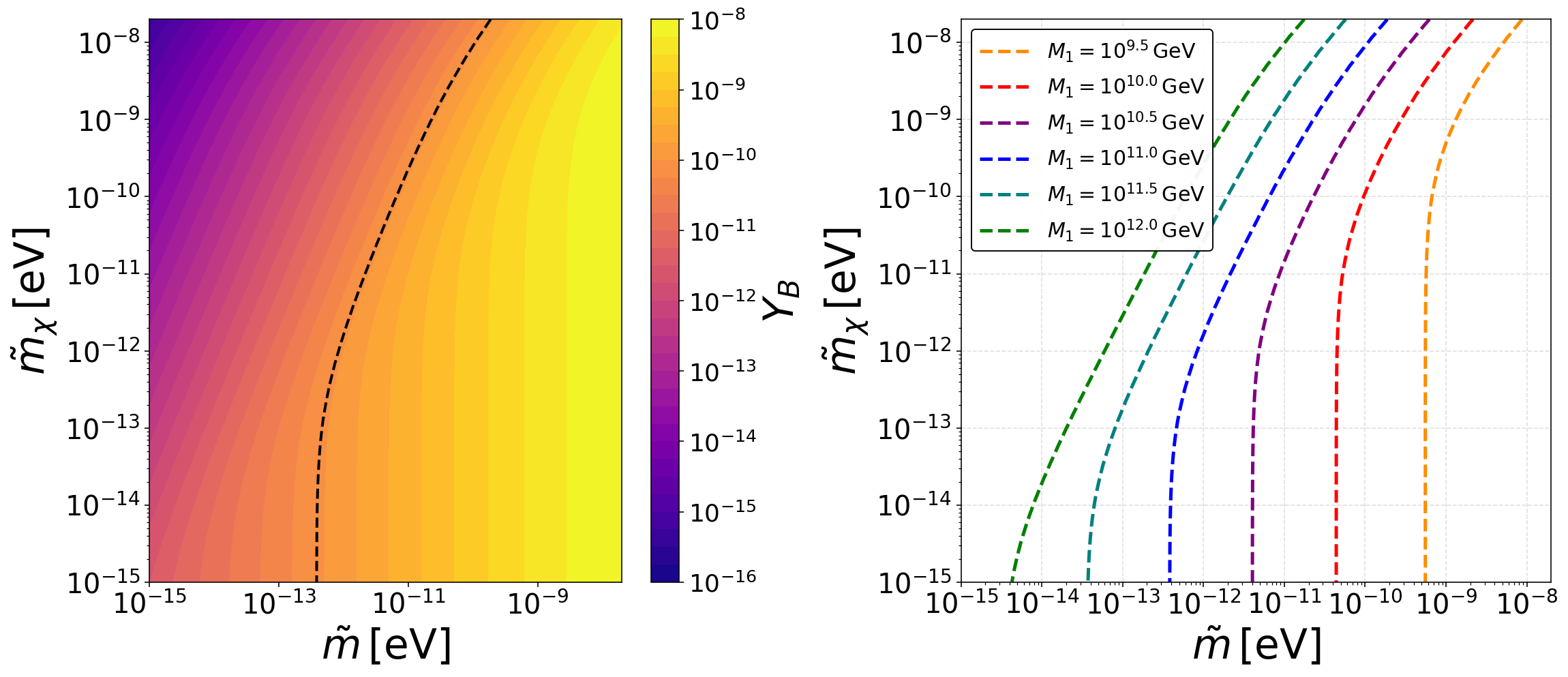} 
\caption{\it \textbf{Left panel:} Scans of baryon asymmetry for $M_1=10^{11}$ GeV, the black dashed line denotes the observed baryon asymmetry, $Y_{B}^{\rm obs}$. \textbf{Right Panel:} $Y_B=Y_B^{\rm obs}$ for various right-handed neutrino masses. Successful leptogenesis is achieved if the parameters are to the right of the $Y_{B}^{\rm obs}$ lines. All points/lines on the plot have a period of early matter domination in their evolution. A very narrow region of detectable parameter space appears at $M_1=10^9\rm GeV$ for both local and global cosmic strings; however, it lies at the extreme edge of the plotted range and is therefore not shown. For $M_1 \geq 10^{10}\rm GeV$ there is no detectable signal. This illustrates that while leptogenesis and dark matter production can be simultaneously realised with relative ease, their experimental accessibility remains severely constrained by the limited detectability of leptogenesis itself.}
    \label{fig:DM lepto}
\end{figure}

We now turn to leptogenesis, which is determined by the lightest right-handed neutrino. Thus, we specify $N \rightarrow N_1$ and correspondingly set the mass as $M \rightarrow M_1$. In this case, the viable parameter space is determined by three quantities: the lightest right-handed neutrino mass $M_{1}$, the corresponding effective neutrino mass $\tilde m$, and the effective dark matter mass parameter $\tilde m_{\chi}$. Together, these parameters uniquely fix the region in which both the observed baryon asymmetry and the correct dark matter abundance can be obtained. We will show that successful dark matter production and baryogenesis can be readily achieved; however, since the testability of leptogenesis is limited to the boundary of the hierarchical regime, the detectability of both mechanisms remains severely constrained. In Fig.~\ref{fig:lepto effective mass scan}, we show the dependence of the baryon asymmetry on the effective masses. The scan ranges from the upper bound required for a period of early matter domination down to the numerically determined minimum effective mass for successful baryogenesis, as obtained in Sec.~\ref{sec:High_scale_lepto}. The baryon asymmetry grows with $\tilde m$ but decreases with $\tilde m_{\chi}$. This behaviour reflects the fact that the branching ratio into Standard Model particles increases with $\tilde m$ while it is suppressed by larger values of $\tilde m_{\chi}$.
We performed a scan over the effective mass parameter space, bounded from below by the condition for successful leptogenesis and from above by the requirement of early matter domination, while varying the right-handed neutrino mass. This allowed us to identify the regions compatible with successful leptogenesis. In Fig.~\ref{fig:DM lepto}, we illustrate the baryon asymmetry obtained for $M_1 = 10^{11}\mathrm{GeV}$ together with the corresponding bounds on the effective mass for different right-handed neutrino masses. The detectability in this scenario is highly limited, primarily due to the restricted testability of leptogenesis. As shown in Fig.~\ref{fig:lepto experiment mass}, only a small region with $M_1<10^{10}\rm\ GeV$ lies within the range of potential experimental sensitivity.
The right-handed neutrino mass \(M_1\) together with the total effective mass parameters \(\tilde m_{\mathrm{tot}} \equiv \tilde m + \tilde m_\chi\) fixes the thermal history of the early Universe: these two parameters determine the onset and termination of the matter–dominated epoch,
\(T_{\mathrm{dom}}(M_1,\tilde m_{\mathrm{tot}})\) and \(T_{\mathrm{end}}(M_1,\tilde m_{\mathrm{tot}})\).
From \(T_{\mathrm{dom}}\) and \(T_{\mathrm{end}}\) one obtains the characteristic gravitational–wave
frequencies \(f_{\mathrm{dom}}\) and \(f_{\mathrm{brk}}\), which set whether the feature lies within the
sensitivity window of future detectors. Simultaneously, the triplet \((M_1,\tilde m,\tilde m_\chi)\) controls leptogenesis, since
\(M_1\) sets the scale while \(\tilde m\) and \(\tilde m_\chi\) fix the decay branching ratios and entropy dilution.
Dark matter production is determined by the two effective mass parameters \(\tilde m\) and
\(\tilde m_\chi\) provided the kinematic condition \(M_1 > M_{\mathrm{DM}}\) is satisfied.
Thus the same parameters that set \(T_{\mathrm{dom}}\) and \(T_{\mathrm{end}}\) and hence
\((f_{\mathrm{dom}}, f_{\mathrm{brk}})\) also govern the viability of leptogenesis and dark matter.\\
These results demonstrate that realizing both dark matter and leptogenesis within the model is straightforward, and that combining dark matter with the experimentally testable parameter space is also easily achieved, as shown in Figure \ref{fig:DM effective mass scan}. By contrast, simultaneously satisfying leptogenesis and experimental testability proves far more restrictive. Consequently, the ability to test the full framework is ultimately limited by the testability of leptogenesis itself.

\medskip

\section{Discussion and Conclusion}
\label{sec:conclusion}

In this work we have investigated the cosmological implications of long-lived heavy seesaw states within the framework of both global and gauged $U(1)_{B-L}$ symmetry breaking. Our analysis has identified a number of novel and testable features correlating right-handed neutrino dynamics, early matter domination, gravitational waves, and the generation of the baryon asymmetry and dark matter.\\
We first demonstrated that a period of early matter domination induced by heavy, long-lived particles can indeed occur. In the context of heavy seesaw states, right-handed neutrino domination arises in both the global and gauged $U(1)_{\rm B-L}$ realizations of the Type-I seesaw mechanism: the global case requires non-thermal production of right-handed neutrinos, whereas the gauged case can be achieved with purely thermal production. Crucially, we show that in the Type-II and Type-III seesaw frameworks, the heavy states cannot induce an early matter-dominated phase, as their interactions prevent an efficient freeze-out. These results hold independently of the light-neutrino mass ordering. By first deriving analytical estimates for the onset temperature and duration of the matter-dominated era, we established their dependence on the right-handed neutrino mass \( M \) and the effective neutrino mass \( \tilde{m}_i \). We then solved the full Boltzmann equations across all relevant parameter regimes and found that these analytical expectations are well captured by simple best-fit relations: the condition for early matter domination reduces to a single requirement on the effective neutrino mass (Eq.~\ref{eq: numeric mtilde}), while the onset temperature and duration of the matter-dominated era are determined solely by \( M \) (Eq.~\ref{eq: Tdom}) and \( \tilde{m}_i \) (Eq.~\ref{eq: MD duration}), respectively. \\
The key consequence of this early matter-dominated phase driven by right-handed neutrinos is its impact on the stochastic gravitational wave background generated by cosmic strings associated with $U(1)_{B-L}$ breaking. The modified expansion history alters the spectral shape of the GWB, providing a potential observational probe of RHN-induced early matter domination. A particularly compelling and unique signature, for a transient, brief matter-dominated era due to these mesta-stable RHN, is a sharp, step-like feature in the GW spectrum, depicting observable kinks in the GW spectrum, one corresponding to the onset of early matter domination and the other to its end (see Fig. \ref{fig:CS benchmark}). We analysed whether such characteristic features in the GW spectrum could be detected in next generation GW detectors such as LISA and ET. We showed that in the context with global cosmic strings, one will be able to probe right-handed neutrino masses across nine orders of magnitude, from $M \sim 10\,\mathrm{GeV}$ up to $M \sim 10^{9}\,\mathrm{GeV}$, with sensitivity down to effective masses of order $\tilde m \sim 10^{-9}\,\mathrm{eV}$. The same for local strings extend this reach considerably, covering right-handed neutrino masses from $M \sim 0.1\,\mathrm{GeV}$ scale to $10^{9}\,\mathrm{GeV}$, while probing neutrino mass parameters as small as $\tilde m \sim 10^{-10}\,\mathrm{eV}$. Symmetry breaking with a larger $v_{\rm B-L}$ leads to an increased amplitude of the gravitational-wave background and shifts its characteristic frequencies to lower values, and is therefore easier to detect the GW signal itself and the characteristic features on top of it (see Figs. \ref{fig:kappaThermal1} and \ref{fig:kappaThermal2}). We presented some analytical results correlating the characteristic frequencies involving the break and knee and the seesaw parameters $\tilde m$, $M$ (see Eqns. (\ref{eq:fbreak}---\ref{eq:fdom} and \ref{eq: Tdom}, \ref{eq: Tend}).) via identifying the start and end temperatures of the period of RHN domination. The results are different for gauged or global $B-L$ extensions. By estimating the SNR for various GW detectors ( \ref{fig:kappaThermal1} and \ref{fig:kappaThermal2}) and combining different sets of GW detectors we found the discovery regions of such features which provide invaluable information and a concrete evidence for a new stage in the cosmological expansion history, enabling us to pin down the start and end of $N$ domination, thereby determining the suppression of the RHN mass scale ($M_1$) compared to the scale of spontaneous symmetry breaking ($v_{\rm B-L}$). The step-by-step pathway connecting measurable features of the gravitational-wave spectrum to the fundamental seesaw parameters is summarised in Fig.~\ref{fig:GWB_inference_chain}. The resulting novel features compatible with observed baryon asymmetry and DM relic, for instance, LISA will be able to detect $M_1 \sim 10^{6}$ GeV. The detectable regions of the parameter space are shown in Figure \ref{fig:Mass experiment bound}, showing that a large region of the parameter space can be probed through gravitational-wave observations. \\
We further explored the implications of this framework for leptogenesis. Considering both thermal and non-thermal initial abundances of right-handed neutrinos, we found that flavour effects are negligible due to the weak-washout regime. The dynamics of leptogenesis are, however, modified by entropy dilution arising from the late decays of the right-handed neutrinos during the matter-dominated era. This framework nevertheless allows analytic bounds to be derived for hierarchical leptogenesis. Near-resonant leptogenesis was also analysed through numerical parameter scans, showing that the edge of the hierarchical regime can be probed via gravitational waves from cosmic strings, while near-resonant leptogenesis lies well within the reach of upcoming experiments. The regions of successful leptogenesis and experimental testability are shown in Figure~\ref{fig:lepto experiment mass}.\\
Finally, we extended the analysis to scenarios where the right-handed neutrinos also decay into a dark sector. In this context, we examined both symmetric and asymmetric dark-matter production, deriving bounds on the viable dark-matter mass consistent with the observed relic density. The dark-matter abundance is likewise affected by entropy dilution from the late decays of the right-handed neutrinos, which modifies the relation between the decay parameters and the final relic density. We showed that the co-genesis of dark matter and the baryon asymmetry is straightforward, although its testability is primarily determined by the leptogenesis constraints. If leptogenesis is relaxed, however, near-future gravitational-wave and collider experiments could readily probe the region of parameter space where the heavy neutrinos predominantly decay into the dark sector. The relevant detectable parameter space for this scenario is shown in Figure~\ref{fig:DM effective mass scan}. The mass for dark matter can range between the Lyman alpha bound $\mathcal{O}(10\ \rm KeV)$ and the right-handed neutrino mass, $M$, which value is then fixed by the two effective neutrino mass parameters $\tilde m,\ \tilde m_\chi$.\\
Once the characteristic features of the GW spectral shapes alluded to in this study are observed, one may look to target additional observations to distinguish between a metastable RHN-dominated pre-BBN era and other forms of early matter domination. Particularly in low-scale ARS leptogenesis\cite{Akhmedov_1998, Drewes_2018, Klaric_2021}  RHN masses of GeV-TeV could be searched in typical Heavy Neutral Lepton Searches (HNL) search experiments (see~\cite{Beacham:2019nyx,Abdullahi:2022jlv,Middleton:2022dio} for current constraints and future experimental sensitivities) at the particle physics laboratories and astrophysical observables. In this manner, we can complement GW searches with laboratory searches in the same BSM parameter space. Nonetheless, such a study is beyond the scope of the present paper. We will explore this in future. Another complementary search for RH neutrino, if they exist, involves experiments such as neutrino-less double beta decay~\cite{Dolinski:2019nrj,Gomez-Cadenas:2011oep} or lepton number violating processes~\cite{Li:2021fvw,Cai:2017mow} and therefore provide us with a myriad of pathways to independently verify the existence of an early RHN-domination era, see Ref. \cite{Borboruah:2025hai} for first steps towards such a complementary study but in the context with inflationary GW. This leads to a unique and exciting opportunity to form synergies between GW observations and laboratory searches.\\
In summary, our results establish a unified framework correlating right-handed-neutrino dynamics, early matter domination, and observable gravitational-wave signals. The underlying \( U(1)_{B-L} \) symmetry simultaneously governs the generation of right-handed-neutrino masses, baryogenesis, dark-matter production, and cosmic-string formation, thereby connecting microphysical physics with cosmological observables. This framework offers a coherent picture in which gravitational-wave observations provide a direct window into the shared origin of neutrino masses, baryogenesis, dark matter, and the early-universe dynamics of the \( U(1)_{B-L} \) sector under one umbrella.

\section*{Acknowledgements}
We thank Pasquale Di Bari for many valuable discussions on leptogenesis and Lekhika Malhotra for comments. The work of SD is supported by the National Natural Science Foundation of China (NNSFC) under grant No. 12150610460.
GW and AS acknowledge the STFC Consolidated Grant ST/X000583/1. AS thanks the University of Southampton School of Physics and Astronomy for the support of a Mayflower PhD scholarship.

\bibliographystyle{apsrev4-1}
\bibliographystyle{utphys}
\bibliography{main}
\end{document}